\documentclass[twocolumn,aps,nofootinbib]{revtex4-2}
\usepackage{graphicx}
\usepackage{epstopdf}
\usepackage{hyperref}
\usepackage{latexsym}
\usepackage{amsmath}
\usepackage{amssymb}

\usepackage[utf8]{inputenc}
\hypersetup{
    colorlinks=true,
    linkcolor=red,
    citecolor=blue,
}
\usepackage{color}
\usepackage[T1]{fontenc}
\usepackage{txfonts}
\usepackage{supertabular}
\usepackage{setspace}
\usepackage{multirow}
\usepackage{makecell}
\usepackage{mathrsfs}

\usepackage[center]{subfigure}

 \newcommand{\bq}{\begin{equation}}
 \newcommand{\eq}{\end{equation}}
 \newcommand{\bqn}{\begin{eqnarray}}
 \newcommand{\eqn}{\end{eqnarray}}

\makeatletter

\newcommand{\Rmnum}[1]{\expandafter\@slowromancap\romannumeral #1@}

\makeatother

\begin{document}
\title{Arm locking in conjunction with time-delay interferometry}
\author{Pan-Pan Wang\textsuperscript{1}}
\author{Wei-Liang Qian\textsuperscript{2,3,4}}\email[E-mail: ]{wlqian@usp.br}
\author{Han-Zhong Wu\textsuperscript{1}}
\author{Yu-Jie Tan\textsuperscript{1}}
\author{Cheng-Gang Shao\textsuperscript{1}}\email[E-mail: ]{cgshao@hust.edu.cn}

\affiliation{$^{1}$ MOE Key Laboratory of Fundamental Physical Quantities Measurement, Hubei Key Laboratory of Gravitation and Quantum Physics, PGMF, and School of Physics, Huazhong University of Science and Technology, Wuhan 430074, China}
\affiliation{$^{2}$ Escola de Engenharia de Lorena, Universidade de S\~ao Paulo, 12602-810, Lorena, SP, Brazil}
\affiliation{$^{3}$ Faculdade de Engenharia de Guaratinguet\'a, Universidade Estadual Paulista, 12516-410, Guaratinguet\'a, SP, Brazil}
\affiliation{$^{4}$ Center for Gravitation and Cosmology, College of Physical Science and Technology, Yangzhou University, Yangzhou 225009, China}

\date{Sept. 11th, 2022}

\begin{abstract}
A crucial challenge to the ongoing endeavor of spaceborne gravitational wave (GW) detection resides in the laser phase noise, typically 7 to 8 orders of magnitude above the inevitable noise.
The arm locking technique was proposed to suppress the noise in pre-stabilized laser beams.
Based on the feedback control theory, it is implemented by appropriate design of the signal routing architecture, particularly the controllers' transfer functions.
Theoretically and experimentally, the technique has been demonstrated to be capable of suppressing the laser phase noise by approximately 2-4 orders of magnitude while taking into account various aspects such as the gain and distribution of nulls in the Bode plot and the laser frequency pulling associated with the Doppler frequency subtraction.
Consequently, the resultant noise floor is composed of the sources attributed to the clock jitter, optical bench motion, test mass fluctuations, shot-noise phase fluctuations at
the photodetectors, whereas the magnitudes of these noises largely remain unchanged during the process.
Besides, the original GW signals are deformed through the arm-locking control loop and therefore bear specific features governed by the associated arm-locking scheme.
Nonetheless, the remaining laser phase noise from the arm-locking feedback routing settles within the capability threshold of the time-delay interferometry (TDI).
In this regard, it is generally understood that the output of arm locking furnishes the input of TDI, through which the residual noise is further reduced to the desired level at a post-processing stage.
In this work, we investigate the specific schemes regarding how the arm locking output is processed further by the TDI algorithm.
Specific forms of the TDI combinations are derived in accordance with suppressed laser phase noise and deformed signals of GW.
To the best of our knowledge, such explicit TDI schemes aiming at the processed signals by the arm locking technique have not been explored in the literature.
Also, we propose a real-time acousto-optic modulation scheme to compensate for the noise due to optical bench motion.
The resultant noise floor, in turn, is primarily composed of those due to the test mass noise and shot noise.
The sensitivity curves are evaluated and indicate that the resulting performance meets the requirement of the LISA detector.
Although the forms of the obtained arm-locking TDI solutions are different from the conventional ones, the response functions, residual noise power spectral densities, and sensitivity curves are found to be identical to their counterparts.
Further implications of the present findings are also addressed.

\end{abstract}

\maketitle


\section{Introduction}\label{section1}

The direct observation of GW offers a powerful tool to extract essential information from astrophysical processes that involve massive astrophysical sources that have significant implications in the fields of fundamental physics and cosmology.
These processes include coalescences of binary black holes and neutron stars, supernovae explosions, and gamma-ray bursts.
Since the first detection of GW reported by the LIGO in 2016~\cite{LIGO-01}, numerous GW events have been encountered~\cite{LIGO-02, LIGO-03,LIGO-04,LIGO-05,LIGO-06,LIGO-07}.
Like the electromagnetic signals, the GW spectrum also covers a wide range in the frequency space. 
Ground-based GW detectors, such as LIGO~\cite{gw-ligo1,gw-ligo2}, VIRGO~\cite{gw-virgo}, and KAGRA~\cite{gw-KAGRA1,gw-KAGRA2}, are more suitable for the detection of high-frequencies (10 Hz-1 kHz) GWs.
At lower frequencies, the detection is mainly hampered by the earth's seismic and gravity-gradient noises.
In order to access the low-frequency band  (0.1 mHz-1 Hz) of the GW spectrum, a few spaceborne GW detector programs have been proposed and are under active development, notably LISA~\cite{gw-lisa1,gw-lisa2}, TianQin~\cite{gw-tianqin}, and TaiJi~\cite{gw-Taiji}.

Characterized by the ultra-narrow linewidth and excellent stability, laser interferometry allows for ultra-precision measurement of phase variation.
At LIGO and future LISA, the GW signals are therefore captured through the interferometry of coherent laser beams in terms of the phase deviation imprinted by the marginally conceivable deformation of the spacetime curvature.
Indeed, the GWs are extremely weak when arriving at the detectors, whose estimated magnitudes correspond to a relative length change $\Delta L/L\sim 10^{-20}{\rm{/Hz}}^{1/2}$ in the science bands~\cite{gw-lisa1}.
Subsequently, it poses tough challenges to further suppress the relevant noise even for the state-of-the-art cavity-stabilized lasers and ultra-stable oscillators.
The most dominant noise for the spaceborne GW detectors originates from the laser phase fluctuations.
While it was not a major problem for the ground-based counterparts, it turns out not feasible to hold a constant-arm constellation for spacecraft following orbits at the scale of $10^8-10^9$ m.
In other words, one cannot straightforwardly establish an equal-arm Michelson interferometer setup in space in order to cancel out the laser phase fluctuations.
Moreover, the clock jitter, optical-bench motion, fiber noise, test-mass vibration, and shot noise significantly affect the measurement performance.
To this end, the time delay interferometry (TDI)~\cite{tdi-01,tdi-02,tdi-03} was proposed to reduce the laser phase noise at the postprocessing stage. 
Specifically, through an appropriate linear combination of the time-delayed data streams, the laser phase noise\footnote{It is also referred to as laser frequency noise in the literature.}, which assumes an arbitrary but given profile, is precisely aligned and canceled out.
Physically, the above process establishes a virtual equal-arm optical interferometer, as the combination can be generated by the geometric TDI approach.
On the other hand, from a computational algebraic geometry perspective, a feasible TDI solution is furnished by the generators of the {\it first module of syzygies}.
In addition to the laser phase noise, the clock noise and the optical bench motion noise can also be handled in the framework of TDI to meet the requirement of LISA.
This is accomplished by introducing additional data streams, particularly the test mass, reference, and sideband interferometric measurements~\cite{tdi-clock4,tdi-clock3,tdi-clock-Wang}, besides the original science data stream.
To date, different schemes of the TDI algorithm have been continuously proposed and investigated for over the past two decades, which can be classified into two generations~\cite{tdi-d77,tdi-d88,tdi-d99,tdi-laser-01,tdi-laser-04,tdi-laser-LISACode}. 
The first generation of TDI combinations is mostly suitable to the scenario of static constellation, namely, the detector's arm length are different but invariant in time~\cite{tdi-01,frame-01-2000,tdi-d55-2001,res-semi--01-2002,tdi-geome-2002,tdi-d22,tdi-laser-06,tdi-laser-LISACode}.
Regarding to the time-variant arm length associated with the spacecraft's orbital motion, the second generation TDI has been developed~\cite{tdi-laser-01,tdi-d99,tdi-d88,tdi-2010-Dhurandhar,tdi-clock4,tdi-filter-s4}.
The improvement is achieved by eliminating the higher-order residuals associated with the relative motion and acceleration of the spacecraft.
Second-generation TDI combinations, such as the Michelson and Sagnac ones, have been extensively explored in the literature.
If one laser link becomes unavailable, the Monitor, Beacon, and Relay combinations will be qualified to the task~\cite{tdi-otto-2015}.
Nonetheless, as discussed above, owing to the spacecraft's orbital motion, the TDI algorithm cannot entirely eliminate the laser phase noise. 
Analytical and numerical studies~\cite{arm-moddual-2009,tdicapability-2009} indicated that the noise of the laser beams post-processed by the TDI must be below a given capability threshold.
In this regard, the arm locking technique was proposed~\cite{arm-theoretically-2003,arm-theoretically-2005,arm-theoretically-2006} to suppress the noise in pre-stabilized laser beams and therefore guarantee TDI's successful implementation.

Based on the feedback control theory, arm locking~\cite{arm-theoretically-2003} was first proposed in 2003 by transferring the stability of the LISA arm length to the laser frequency.
In particular, the approach is implemented by appropriate design of the signal routing architecture, particularly the controllers' transfer functions.
The essential aspects of the technique are associated with the amplitudes and nulls in the Bode plot and the laser frequency pulling associated with the Doppler frequency subtraction.
In its barebone form, the arm locking scheme works for a single-arm interferometry~\cite{arm-theoretically-2003}. 
In the frequency space, the resulting laser phase noise is roughly suppressed by a factor of the gain, the amplitude of the controller's transfer function.
However, in the meantime, a series of nulls in the transfer function, which correspond to twice the time delay between the two spacecraft, will also consequently be populated in the science band, leading to potentially significant noise amplification.
Regarding this limitation, dual-arm locking was developed~\cite{arm-dual-2008}.
It was shown that the first null could be shifted out of the LISA science band by introducing an appropriate feedback control by combining the signals from two interferometer arms.
Despite the substantial noise reduction, the technique suffered more significantly from the frequency pulling, related to the error in the Doppler frequency estimate. 
Subsequently, a hybrid combination between the common and dual-arm locking was introduced~\cite{arm-moddual-2009}.
Such a scheme inherits the delimited laser frequency pulling and low-frequency noise coupling of common arm locking while retaining the advantage of dual-arm locking. 
More recently, the compatibility between the cavity stabilization and the arm locking was also discussed, and a new type of arm-locking sensor integrated with the Pound-Dever-Hall (PDH) error signals has been proposed~\cite{arm-the-2022}.
In this scenario, the noise is feedback controlled by the hybrid combination between the arm locking and PDH locking loops, and no additional hardware is required.
In this regard, the laser phase noise can be effectively suppressed by 2-4 orders of magnitude and satisfactorily meet the requirement of TDI capacity. 
The resultant noise sources mainly consist of the clock jitter at lower frequencies, optical bench motion in the higher frequency band, sitting on top of the test mass fluctuations, and shot noise, whereas the magnitudes of these noises largely remain unchanged during the laser phase noise suppressing process.
As discussed above, some of these noises can also be dealt with within the framework of the TDI algorithm.
It is worth noting that the original GW signals are deformed through the arm-locking control loops that, in turn, bear specific features governed by the associated arm-locking scheme.
Nonetheless, as a result, the output of arm locking furnishes the input of TDI, through which the remaining noise can be further suppressed to the desired level. 

While the conjunction of the two methods, namely, arm locking and TDI, is understood to be a favorable choice in practice, it is arguable that such a union gives birth to a few novel aspects that deserve further scrutinization.
For instance, it is not straightforward how the cavity-stabilized laser locked to the constellation baselines can be fed to the Monitor, Beacon, and Relay combinations.
This is because, in the arm-locking closed loop, the laser in the distant spacecraft is inherently phase-locked to the incoming beam, emanated from the so-called master laser using carrier-carrier measurement as the error signal.
The latter removes certain degrees of freedom regarding the phase between different laser beams, which might be exploited at the post-process stage to cancel out laser phase noise.
Moreover, as mentioned, the forms of various types of noise, and inclusively the laser phase noise, are distorted as a result of the arm locking process.
Therefore, it is not clear whether the existing TDI schemes can be readily applied to the output signals of arm locking.
Indeed, it is expected that the resulting TDI combinations will exhibit new features, and the forms of the resultant residual noise will also be modified accordingly.
To the best of our knowledge, the TDI formulation in conjunction with arm locking has not been explicitly derived in the literature, and the present study is motivated by the above considerations.
As a first attempt, the present paper will refrain from dealing with the second-generation TDI.
As discussed in Appendix~\ref{appd2}, it can be justified since the presence of the arm locking provides an additional 2-4 orders of magnitude in laser phase noise suppression, and therefore the use of first-generation TDI suffices for the required sensitivity leading to possible GW detection.
In this work, we examine the compatibility between arm locking and TDI by deriving specific forms of the TDI combinations.
Also, we elaborate on the cancelation schemes of other types of noise.
We suggest a real-time compensation scheme for the optical bench displacement in the case of the optical bench vibration noise.
Subsequently, the magnitude of the optical bench noise will be suppressed to the level of the test mass one, which is more insignificant by several orders of magnitude.
As a result, the noise floor is primarily determined by the test mass noise and shot noise. 
We evaluate the noise power spectral densities (PSDs) and resulting sensitivity curves while discussing the feasibility of the present scheme by comparing the obtained results with existing ones.

The remainder of the paper is organized as follows.
Before diving into any specific expressions, in Sec.~\ref{section1.5}, we present an overview of our results by comparing them against the standard TDI formulation.
Sec.~\ref{section2} gives a brief account of the experimental setup, the schemes of laser phase noise suppression in the frequency domain for spaceborne GW detectors, and the notations adopted in the present work.
In Sec.~\ref{section3}, we elaborate on the basics of the TDI algorithm, including its essential elements such as various data streams, different types of combinations, and the detector sensitivity curve.
In Sec.~\ref{section4}, we discuss the arm locking technique.
For the present study, we elaborate on a specific real-time compensation scheme based on the acousto-optic modulation (AOM) to effectively suppress the optical bench vibration noise.
Also, the forms of the resultant GW signal and noise are derived primarily in terms of the controller's transfer functions.
Using the relevant observables derived in the preceding section as the input, Sec.~\ref{section5} presents our main results by deriving the relevant TDI combinations and the corresponding sensitivity curve.
The last section, Sec.~\ref{section6}, is devoted to further discussions and concluding remarks.
Further discussions regarding the magnitude of noise suppression of various methods and the detailed derivations of some of the results are relegated to the Appendices of the paper.

\section{Summary of the results}\label{section1.5}

We first briefly summarize the main results obtained in this study by pinpointing them in the subsequential sections.
A standard TDI solution typically possesses the form of Eq.~\eqref{tdi}, where the summation is carried out for the observables defined by Eqs.~\eqref{postpre} and~\eqref{postpre1}. 
The coefficients that define the specific linear combination of observables are essentially polynomials of the frequency-domain time-delay operators Eq.~\eqref{delayop}.
For a valid combination furnished by a specific set of coefficients, such as those given by Eqs.~\eqref{tdipoly} in the case of the Michelson combination, the laser phase noise cancels out in the resultant expression.
Therefore, the results of the present work are also given in the form of Eq.~\eqref{tdi}.
Nonetheless, there are a few notable differences.
First, the relevant observables are modified due to the arm locking procedure.
They are generally different from the standard TDI variables and dependent on the specific arm-locking scheme, such as the controllers' transfer functions. 
In this paper, they are given by Eqs.~\eqref{finallyeta1} to~\eqref{appet3pie} in Sec.~\ref{section5.1}, simplied from the forms Eqs.~\eqref{PHI1} to~\eqref{eta22222} derived in Sec.~\ref{section4.2}.
Subsequently, the TDI solution should be derived accordingly.
The resulting TDI coefficients, again taking the Michelson combination as an example, are presented by Eqs.~\eqref{coffXX}.
Although the algebraic form of Eqs.~\eqref{coffXX} is distinct from those of Eqs.~\eqref{tdipoly}, the response functions, residual noise power spectral densities, and sensitivity curves are found to be identical to their counterparts.
We delegate the detailed discussions to the following sections.

\section{Experimental layout, conventions, and laser phase noise suppresion schemes}\label{section2}

\begin{widetext}

\begin{figure}[!t]
\includegraphics[width=0.80\textwidth]{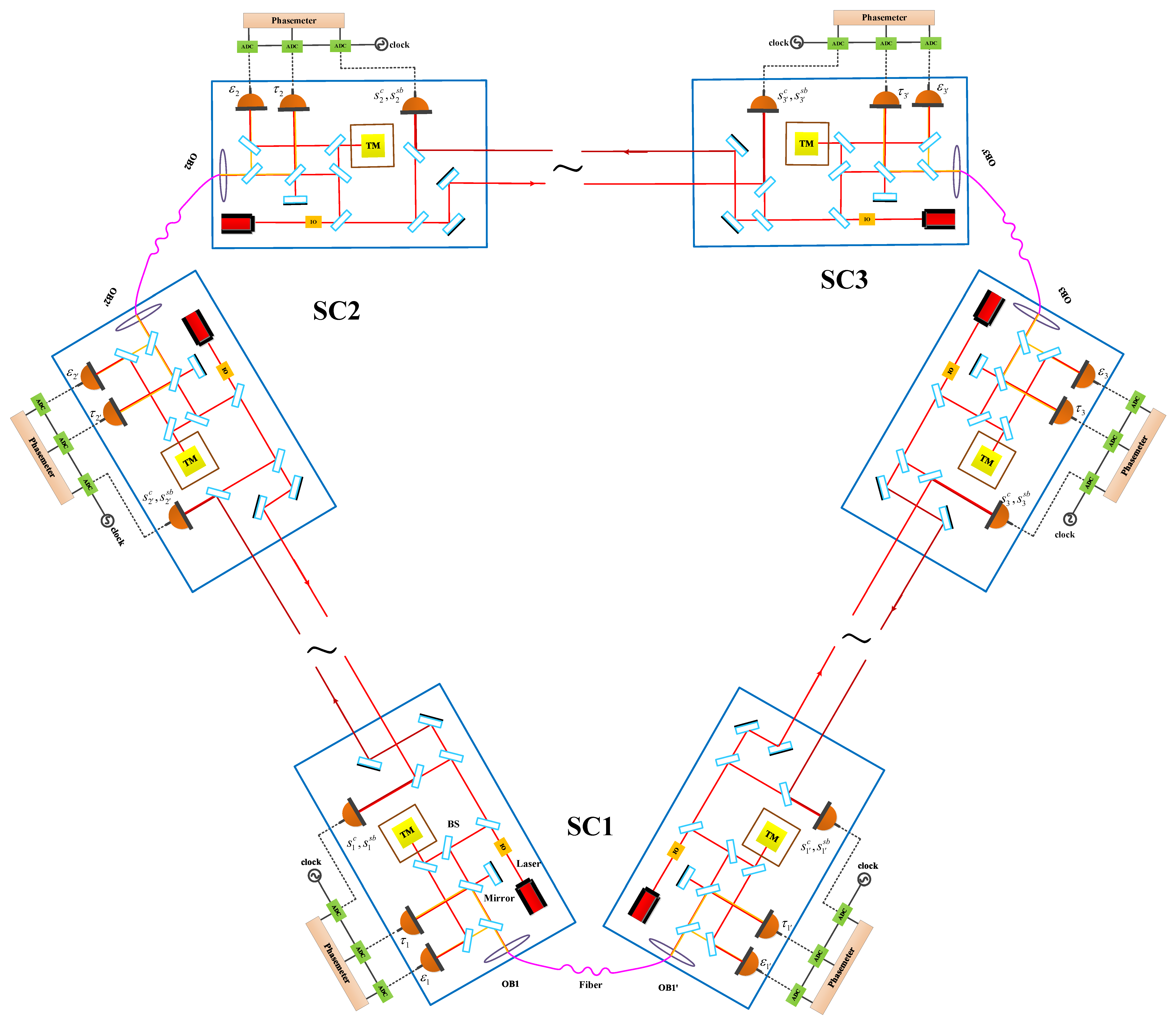}
\caption{\label{fig1} 
Optical and electrical setup of the spacecraft constellation. SC: spacecraft; OB: optical bench; BS: beam splitter; PD: photodetector; IO:isolator; TM:test mass.}
\end{figure}
\end{widetext}

For LISA and TianQin alike, a spaceborne GW detector consists of an almost equilateral enormous triangle formed by three spacecraft, as shown in Fig.~\ref{fig1}, whose baseline is about $10^8 - 10^9$ m.
Due to the orbital motion, the arm length varies by about $1\%$ during the course of a year. 
The onboard micro-Newton thruster controls the posture of the spacecraft.
The laser and test mass are installed on the ultralow expansion optical bench.
An individual spacecraft possesses two optical benches connected via a piece of backlink fiber.
Therefore, there are a total of six lasers for the triangular constellation setup. 
As shown in Fig.~\ref{fig1}, the spacecraft are labeled 1, 2, and 3 in a clockwise direction.
The two identical optical benches located on spacecraft $i$ are denoted by $i$ and $i'$, where $i = 1, 2, 3$.
We use the single-index convention in the present paper, consistent with Ref.~\cite{tdi-03}. 
The {\it Nd: YAG} laser at 1064 nm is typically chosen to be installed in the spaceborne GW detector because of its high power density, excellent beam quality, narrow linewidth, and preferable optical-to-optical efficiency.
Through some specific scheme, the phases of all the lasers are offset and locked per the master laser.
The latter is prestabilized by the Fabry-Perot cavity~\cite{FP-1983, FP-Thorpe-2008}.
The arm locking is employed to suppress the laser phase noise in the prestabilized laser beams below the TDI capacity.
Subsequently, at the post-process stage, the TDI further reduces various types of noise to extract from the interferometry beatnotes the essential information on the GW.
Generally, the laser phase noise of the free-running Nd:YAG laser can reach the level~\cite{require-tdi-laser-2009}
\begin{align}\label{free}
\delta v_{\rm{free}} \sim 3\times 10^4\frac{{\rm{1Hz}}}{f}\frac{\rm{Hz}}{\sqrt {\rm{Hz}}} ,
\end{align}
where $f$ is the frequency. 
By using the PDH technique, the laser source is stabilized to a high-finesse Fabry-Perot cavity, and the noise can be improved to~\cite{require-tdi-laser-2009,armnewdesign-2022}
\begin{align}\label{pdh}
\delta v_{\rm{PDH}} \sim 30 \times \sqrt {1 + {{\left( {\frac{{2.8{\rm{mHz}}}}{f}} \right)}^4}} \frac{{{\rm{Hz}}}}{{\sqrt {{\rm{Hz}}} }},
\end{align}
which is still 7 orders of magnitude above the requirement of LISA, i.e., about $10^{-6 }\rm{Hz/Hz}^{1/2}$, or the relative value of $10^{-20}\rm{/Hz^{1/2}}$.
As mentioned, arm locking and time delay interferometry techniques have been proposed to overcome this challenge.

\begin{figure}[!t]
\includegraphics[width=0.40\textwidth]{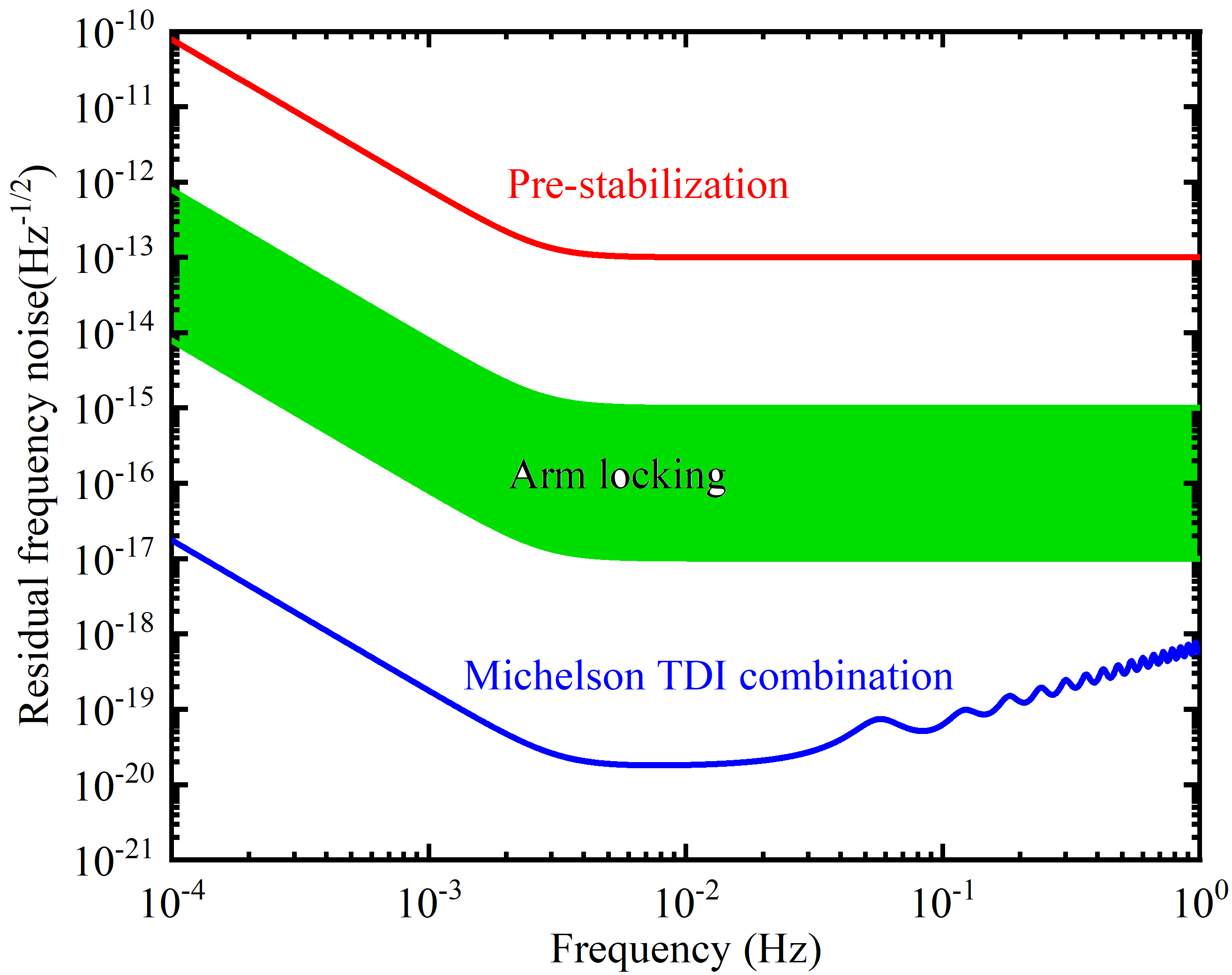}
\caption{\label{fig2} Example of one way to achieve the required frequency stability using pre-stabilization, arm locking and TDI.}
\end{figure}

The arm locking technique utilizes hard wired onboard feedback signal locking to suppress the laser phase noise, and the resulting noise level is expected to be~\cite{arm-moddual-2009}
\begin{align}\label{arm}
\delta v_{\rm{Arm}}\sim [ 0.3,0.003] \times \sqrt {1 + {{\left( {\frac{{2.8{\rm{mHz}}}}{f}} \right)}^4}} \frac{{{\rm{Hz}}}}{{\sqrt {{\rm{Hz}}} }}.
\end{align}
Moreover, as a post-processing algorithm, TDI is utilized to reduce further the laser phase noise to satisfy LISA's requirement where the noise floor primarily consists of the test mass noise and the shot noise.
The frequency-control strategy is schematically shown in Fig.~\ref{fig2}.
The magnitude of the residual phase noise after the prestabilization, arm locking, and TDI processing stages are indicated by the solid red curve, green band, and solid blue curve, respectively.
As an illustration, in Fig.~\ref{fig2}, the resulting noise floor of the Michelson combination is depicted, whose order of magnitude is similar primarily when compared with other feasible TDI combinations.

Both the noise cancellation techniques involve an essential mathematical operation: the time delay.
It is denoted by the time-delay operator ${\cal D}_i$ along the arm lengths $i$ where the prime indicates the inverse cyclic direction.
When applying to the data stream $x(t)$, it gives
\begin{align}\label{delayop}
{{\cal D}_i}x(t) =& x\left(t - {\tau_i}(t)\right),\\\notag
{{\cal D}_j}{{\cal D}_i}x(t) =& {{\cal D}_j}x\left(t - {\tau_i}(t)\right) = x\left(t - {\tau_i}(t - {\tau_j}) - {\tau_j}(t)\right),
\end{align}
where the amount of delay $\tau_i=\frac{L_i}{c}$ or $\tau_j=\frac{L_j}{c}$ is the light travel time between the spacecraft, and $c$ is the speed of light. 
Successive applications of the time-delay operators can be simplified to
\begin{align}
{{\cal D}_j}{{\cal D}_i}x(t) ={{\cal D}_{ji}}x(t).
\end{align}

For convenience, the problem is most elaborated in the frequency domain by employing either Fourier transformation, often adopted for the TDI algorithm, or Laplace transform, utilized for the arm locking formulation largely inherited from the analysis based on the control theory.
In both cases, the transformed time-delay operator possess a more compact behavior as a multiplier.
For the Fourier transform, it becomes
\begin{align}\label{DiFourier}
{\cal D}_i \to e^{-i\omega \tau_i}
\end{align}
in the frequency $\omega$-domain.
For the Laplace transform, it reads
\begin{align}\label{DiLaplace}
{\cal D}_i \to e^{-s \tau_i}
\end{align}
in the $s$-domain.
Regarding the mathematical context the above analyses apply, the two transforms are related by 
\begin{align}\label{FourierXLaplace}
s = i\omega ,
\end{align}
and mostly equivalent. 
Although utilizing two different sets of notations for the delay operators seems somewhat redundant, it brings clarity when referred to with existing literature.
Therefore, in the remainder of the paper, we will adopt the notions associated with the Fourier transform when discussing the TDI algorithm, use those of the Laplace transform when dealing with the arm locking formulation, and, if necessary, address the difference explicitly when a comparison is attempted.

\section{TDI without arm locking}\label{section3}

Before elaborating on the input data streams in the context of arm locking feedback optical routing, it is instructive to give a brief account of the TDI algorithm.
In what follows, we first enumerate the relevant data streams for traditional TDI.
Then, we elaborate on the TDI combinations obtained by canceling out the laser phase noise.

\subsection{The data streams and intermediate variables}

As discussed in Sec.~\ref{section2} and shown in Fig.~\ref{fig1}, the interferometric measurements consists of the science data streams $s_i$ and $s_{i'}$, test mass data streams ${\varepsilon _i}$ and ${\varepsilon _{i'}}$, and reference data streams $r_i$ and $r_{i'}$.
To be specific, the three measurements from optical bench $i$ are
\begin{subequations}
\begin{align}
s_i^c =& {h_i} + {{\cal D}_{i - 1}}{\phi _{(i + 1)'}} - {\phi _i}+ N_i^{S}\notag\\
 +& 2\pi {\nu _{(i + 1)'}}\left[ {{\vec n_{i - 1}} \cdot {{\cal D}_{i - 1}}{\vec \Delta _{(i + 1)'}} + {\vec n_{(i - 1)'}} \cdot {\vec \Delta _i}} \right],\label{s1}\\
{\varepsilon _i} =&{\phi _{i'}}\! -\! {\phi _i} \!+\! {\mu _{i'}} \!-\! 4\pi {\nu _{i'}}\left[ {{\vec n_{(i - 1)'}} \cdot {\vec \delta _i} \!-\! {\vec n_{(i - 1)'}} \cdot {\vec \Delta _i}} \right],\label{v1}\\
{r _i}=&{\phi _{i'}} - {\phi _i} + {\mu _{i'}} ,\label{t1}
\end{align}
\end{subequations}
and the three measurements from optical bench $i'$ are
\begin{subequations}
\begin{align}
s_{i'}^c =& {h_{i'}} + {{\cal D}_{(i + 1)'}}{\phi _{i - 1}} - {\phi _{i'}} + N_{i'}^{S}\notag\\
+& 2\pi {\nu _{i - 1}}\left[ {{\vec n_{i + 1}} \cdot {\vec \Delta _{i'}} + {\vec n_{(i + 1)'}} \cdot {{\cal D}_{(i + 1)'}}{\vec \Delta _{i - 1}}} \right],\label{sp1}\\
{\varepsilon _{i'}} =& {\phi _i} \!-\! {\phi _{i'}} \!+\! {\mu _i} \!-\! 4\pi {\nu _i}\left( {{\vec n_{i + 1}} \cdot {\vec \delta _{i'}} \!-\! {\vec n_{i + 1}} \cdot {\vec \Delta _{i'}}} \right),\label{vp1}\\
{r _{i'}} =& {\phi _i} - {\phi _{i'}} + {\mu _i},\label{tp1}
\end{align}
\end{subequations}
where $h$ corresponds to the GW signal, $\phi$ gives to the laser phase noise, ${\vec \Delta}$ are the optical bench motion noise, ${\vec \delta}$ represents the test mass noise,
${N^{S}}$ indicates the shot noise, and ${\mu}$ is the fiber noise.
Only the science data streams $s_i$ bear the information on the GW signals, test mass $\varepsilon_i$ and reference $r_i$ data streams are therefore auxiliary, introduced merely aiming to remove various noises in the science data stream.
The optical bench motion noise can be removed systematically in the framework of TDI at the post-processing stage or by a hard-wired real-time compensation scheme to be discussed further in Sec~\ref{section4.2}.
For the remaining noise, the test mass noise and the shot noise are considered mainly inevitable, thus furnishing the noise floor of the optical interferometer.
In this work, we will not explicitly address the clock jitter noise, which can also be canceled out by introducing the sideband modulation~\cite{tdi-clock4,tdi-clock-08,tdi-clock-Wang} or laser frequency combs~\cite{tdi-clock5,tdi-clock-09}. 
We relegate further discussions regarding the clock jitter noise to the last section of the paper.

A general approach~\cite{tdi-03} is to first algebraically eliminate the primed laser phase noise and optical bench motion noise, from which one finds
\begin{subequations}
\begin{align}
{\eta _i}=&h_i+{{\cal D}_{i - 1}}{\phi_{i + 1}} \!-\! {\phi_i}\!+\!N_{i}^{S}\!-\!\frac{1}{2}{{\cal D}_{i - 1}}\left[ {{\mu _{(i + 1)'}} \!-\! {\mu _{i + 1}}} \right]
\notag\\
+& 2\pi {\nu_{(i + 1)'}}\left[ {\vec n_{(i - 1)}}\cdot{{{\cal D}_{(i - 1)}}{\vec \delta _{(i + 1)'}}\! +\! {\vec n_{(i - 1)'}}\cdot{\vec \delta _i}} \right],\label{postpre}\\
{\eta _{i'}}=&h_{i'}+{{\cal D}_{(i + 1)'}}{\phi_{i - 1}}\! -\! {\phi_i}\!+\!N_{i'}^{S} \!-\!\frac{1}{2}({\mu _i}\! -\! {\mu _{i'}})
\notag\\
+&2\pi {\nu_{(i - 1)}}\left[{\vec n_{(i + 1)}}\cdot {{\vec \delta _{i'}} \!+\! {\vec n_{(i + 1)'}}\cdot{{\cal D}_{(i + 1)'}}{\vec \delta _{\left( {i - 1} \right)}}} \right].\label{postpre1}
\end{align}
\end{subequations}
The observables $\eta_i,\eta_{i'}$ are generally used in the literature to derive the relevant TDI combinations, where the remaining laser phase noise is canceled out.
It is noted that the resultant test mass noise in Eqs.~\eqref{postpre}-\eqref{postpre1} possesses identical forms as the optical bench motion noise in the original science data given by Eqs.~\eqref{s1} and~\eqref{sp1}.

\subsection{TDI combinations}

The general form of a TDI combination can be expressed as
\begin{align}\label{tdi}
{\rm{TDI = }}\sum\limits_{i = 1,2,3} {({P_i}{\eta _i} + {P_{i'}}{\eta _{i'}})}.
\end{align}
where the coefficients $P_i$ and $P_{i'}$ are the polynomials of the time delay operators in the frequency domain, Eq.~\eqref{DiFourier}.

By applying a given TDI combination of the form Eq.~\eqref{tdi} to the observables Eqs.~\eqref{postpre} and~\eqref{postpre1}, the resulting GW signal reads
\begin{align}\label{tdih}
	{\rm{TDI}}^h = \sum\limits_{i = 1,2,3} {({P_i}{h _i} + {P_{i'}}{h _{i'}})} .
\end{align}
We note the same coefficients in Eq.~\eqref{tdih} and Eq.~\eqref{tdi} are owing to the one-to-one mapping between GW stess and laser noise in the observables Eqs.~\eqref{postpre} and~\eqref{postpre1}.
Also, the residual noise, primarily consisting of the test mass and shot noise, is given by
\begin{align}\label{tdidetlashot}
	{\rm{TDI}}^{\delta+N^S} = &\sum\limits_{i = 1,2,3} {\left[ \begin{array}{l}
			\left( {2\pi {v_{(i + 1)'}}{P_i} + 2\pi {v_i}{P_{{{\left( {i + 1} \right)}^\prime }}}{{\cal D}_{(i - 1)'}}} \right){\vec n_{(i - 1)'}}\cdot{{\vec \delta }_i}\\
			+ \left( {2\pi {v_{i'}}{P_{i - 1}}{{\cal D}_{i + 1}} + 2\pi {v_{i - 1}}{P_{i'}}} \right){{\vec n}_{i + 1}}\cdot{{\vec \delta }_{i'}}
		\end{array} \right]}\\\notag
	+&\sum\limits_{i = 1,2,3} {({P_i}N_{i}^{S} + {P_{i'}}N_{i'}^{S})}.
\end{align}

\begin{figure}[!t]
\includegraphics[width=0.25\textwidth]{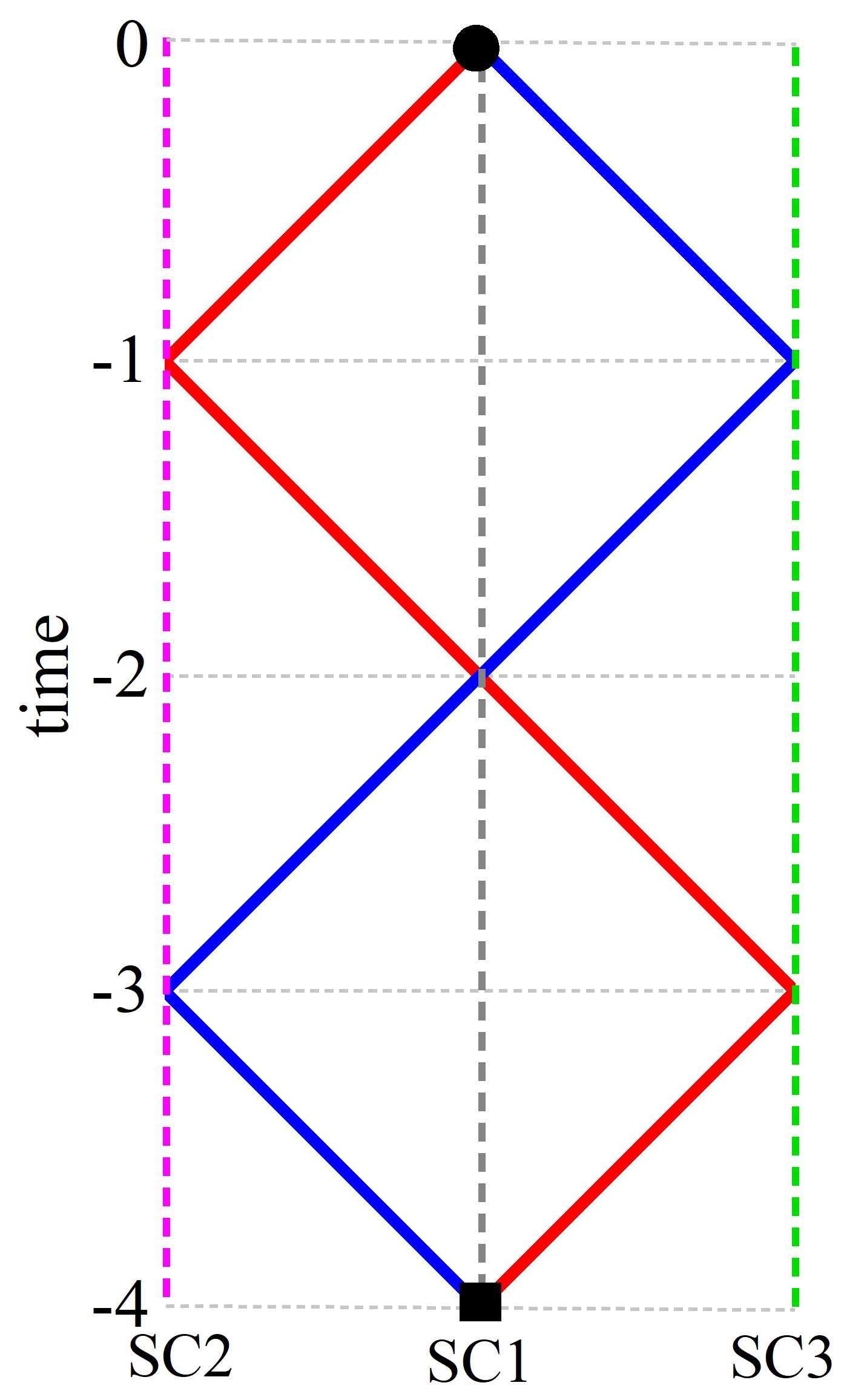}
\caption{\label{fig3} Geometric representation of first generation Michelson TDI combination. 
The red and blue solid line segments correspond to the two synthesized routes. 
The black square and black dot indicate the initial and terminal grids of the two routes, and the reference time $t=0$ is attributed to spacecraft (SC1).}
\end{figure}

As an illustration, we consider the Michelson combination ${X_1}$ shown in Fig.~\ref{fig3}, whose coefficients are given by
\begin{align}\label{tdipoly}
{P_1} =& ({\cal D}_{2'2} - 1),{P_2} = 0,{P_3} = \left( {{\cal D}_{2'} - {\cal D}_{33'2'}} \right);\\\notag
{P_{1'}} =& (1 - {\cal D}_{33'}),{P_{2'}} = ({\cal D}_{2'23} - {\cal D}_3),{P_{3'}} = 0.
\end{align}
To be specific, by Fourier transform, we have
\begin{align}\label{michfour}
{X_1}(\omega ) = & -\! \left( {1 - {e^{ - i\omega ({\tau _2} + {\tau _{2'}})}}} \right){\eta _1}\left( \omega  \right)\! +\! \left( {{e^{ - i\omega {\tau _{2'}}}}\! -\! {e^{ - i\omega ({\tau _3} + {\tau _{3'}} \!+\! {\tau _{2'}})}}} \right){\eta _3}\left( \omega  \right)\notag\\
+ &\left( {1 - {e^{ - i\omega ({\tau _3} + {\tau _{3'}})}}} \right){\eta _{1'}}\left( \omega  \right) - \left( {{e^{ - i\omega {\tau _3}}} - {e^{ - i\omega ({\tau _{2'}} + {\tau _2} + {\tau _3})}}} \right){\eta _{2'}}\left( \omega  \right),
\end{align}
in the frequency domain, where $\omega=2\pi f$.

In the first generation TDI framework, one does not distinguish between the two cyclic directions of a given arm length.
By assuming that ${\tau _1} = {\tau _{1'}}$, ${\tau _2} = {\tau _{2'}}$, and ${\tau _3} = {\tau _{3'}}$, Eq.~\eqref{michfour} can be rewritten as
\begin{align}\label{michfourup}
{X_1}(\omega ) =&  - \left( {1 - {e^{ - i\omega 2{\tau _2}}}} \right){\eta _1}\left( \omega  \right) + \left( {{e^{ - i\omega {\tau _{2}}}} - {e^{ - i\omega (2{\tau _3} + {\tau _2})}}} \right){\eta _3}\left( \omega  \right)\notag\\
+& \left( {1 - {e^{ - i\omega 2{\tau _3}}}} \right){\eta _{1'}}\left( \omega  \right) - \left( {{e^{ - i\omega {\tau _3}}} - {e^{ - i\omega (2{\tau _2} + {\tau _3})}}} \right){\eta _{2'}}\left( \omega  \right).
\end{align}
Subsequently, the residual noise PSD~\cite{response-full-03,response-full-04} in the Michelson combination is found to be
\begin{align}\label{xben}
N(u)_X =16{\sin ^2}u\left[ {\left( {3 + \cos 2u} \right)\frac{{s_a^2{L^2}}}{{{u^2}{c^4}}} + \frac{{{u^2}s_x^2}}{{{L^2}}}} \right] ,
\end{align}
where $u =\frac{\omega L}{c}= \frac{{2\pi fL}}{c}$, $s_x$ is the PSD of the displacement noise, and $s_a$ is that of the acceleration noise.
For LISA detector~\cite{gw-lisa1},
\begin{align}\label{sasx}
{s_a} =& 3 \times {10^{ - 15}}\frac{\rm {m/s^2}}{{\sqrt {{\rm{Hz}}} }}, \\\notag
{s_x} =& 15 \times {10^{ - 12}}\frac{{\rm{m}}}{{\sqrt {{\rm{Hz}}} }}. 
\end{align}
On the other hand, the signal's response function~\cite{response-full-03} reads
\begin{align}\label{responX}
R\left( u \right)_X = \frac{2}{3}{\sin ^2}u\left\{ \begin{array}{l}
5 + \cos 2u\left( {1 - 18\log \frac{4}{3}} \right) + 12\log 2\\
 + \frac{{3\left( { - 7\sin u + 2\sin 2u} \right)}}{u} - \frac{{3\cos u\left( { - 5 + 8\cos u} \right)}}{{{u^2}}}\\
 + \frac{{3\left( { - 5\sin u + 4\sin 2u} \right)}}{{{u^3}}} + 12\left( {{\mathop{\rm Ci}\nolimits} u - {\rm{Ci}}2u} \right)\\
 - 18\cos 2u\left( {{\rm{Ci}}u - 2{\rm{Ci2}}u + {\rm{Ci3}}u} \right)\\
 - 18\sin 2u\left( {{\rm{Si}}u - 2{\rm{Si2}}u + {\rm{Si3}}u} \right)
\end{array} \right\} ,
\end{align}
where SinIntegral ${\rm{Si}}(z) = \int_0^z {\sin t/tdt}$ and CosIntegral ${\rm{Ci}}(z) =  - \int_z^\infty  {\cos t/tdt}$.
Subsequently, the {\it strain} sensitivity curve is governed by the ratio between the signal and noise, which is found to be
\begin{align}\label{requ}
S(u) = \sqrt{\frac{N(u)}{\frac{2}{5}R(u)}} ,
\end{align}
where the coefficient $2/5$ was evaluated for tensorial mode in the framework of General Relativity.
It originates from averaging the orbit inclination for the source from which the GWs are emanated, with respect to the observer's line of sight. 

\section{Arm locking scheme}\label{section4}

\subsection{Principle of arm locking}\label{section4.1}

\begin{figure}[!t]
\includegraphics[width=0.50\textwidth]{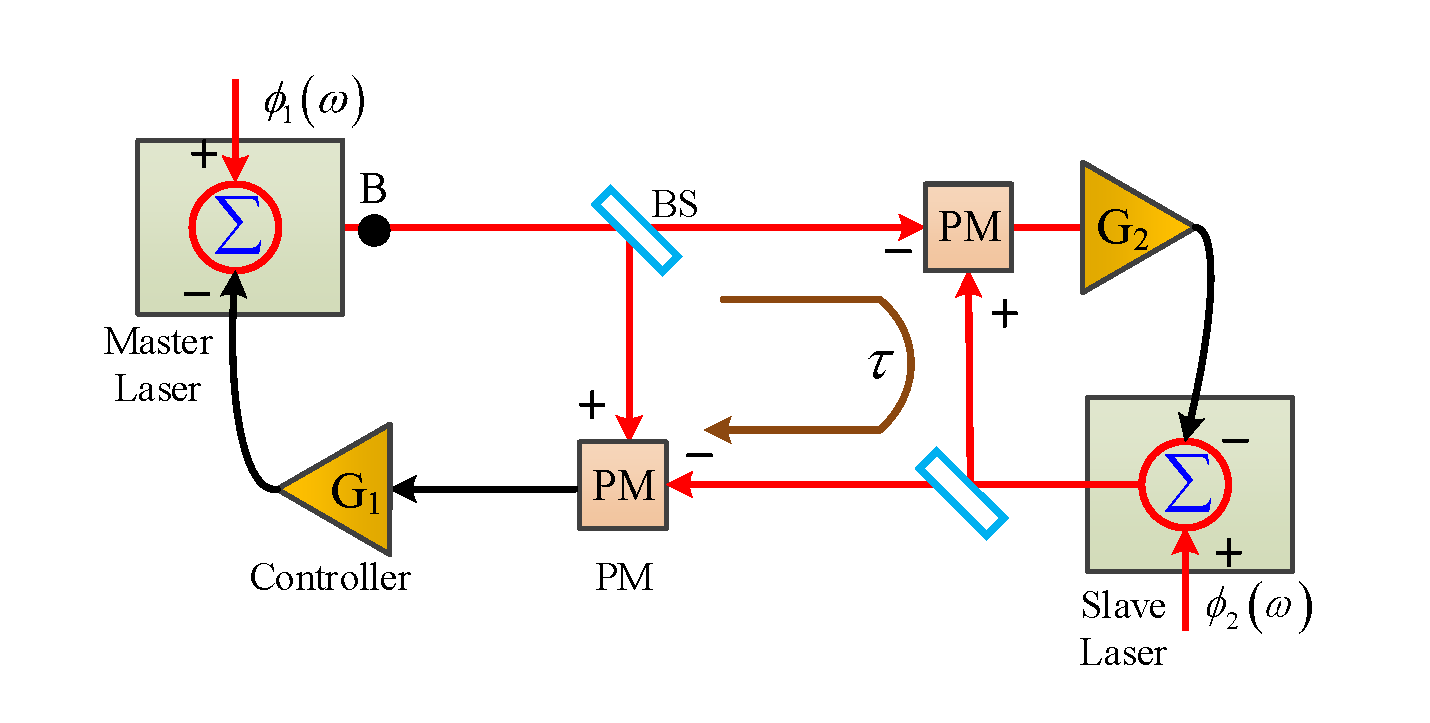}
\caption{\label{fig4}
A schematic layout of arm locking route design with a single arm. 
PM: phasemeter; BS: beam splitter}
\end{figure}

In this section, we first briefly review the principle of arm locking by considering the simplest scenario, which involves only a single arm~\cite{arm-theoretically-2003}.
As shown in Fig.~\ref{fig4}, the light emanated from the master laser is split into two parts.
The signal travels along the arm to reach the distant spacecraft, where the slave laser's phase is locked to the incoming beam and then sends the beam back to the master spacecraft.
The interference between the two laser beams is implemented by a negatively feedback, and the output of the phasemeter can be expressed as
\begin{align}\label{armsig}
	\phi _{\rm{PM}}(t) = \phi \left( t \right) - {\cal D}\phi \left( {t } \right) = \phi \left( t \right) - \phi \left( {t - \tau } \right).
\end{align}
By using Laplace transform Eq.~\eqref{DiLaplace}, the above expression Eq.~\eqref{armsig} can be rewritten as
\begin{align}\label{phipm}
	\phi _{\rm{PM}}(s ) = \phi \left( s  \right) - \phi \left( s  \right)e^{ - s \tau } = T(s) \phi \left( s  \right)
\end{align}
in the $s$-domain, where
\begin{align}\label{fourarmsig}
T(s) = 1 - {e^{ - s \tau }} 
\end{align}
is the transfer function.

In practice, as shown in Fig.~\ref{fig4}, two controllers, $G_1$ and $G_2$, are inserted into the optical feedback route to suppress the laser phase noise.
For instance, the effective transfer function Eq.~\eqref{fourarmsig} for the subloop that involves the controller $G_2$ is subsequently modified to read
\begin{align}\label{TrnasferF1}
T_1(s) = 1 - \frac{G_2(s)}{1+G_2(s)}{e^{ - s \tau }}. 
\end{align}
We note that the subscript on the l.h.s. of the equation is ``1'' since this transfer function is applied to the optical route involving the master laser and controller, both denoted by ``1''.

The resultant close-loop signal at point B is found to be~\cite{arm-theoretically-2003}
\begin{align}\label{armb}
{\phi _B}\left( s  \right) = \frac{{{\phi _1}\left( s \right)}}{{1 + {L_1}\left( s \right)}} + \frac{{{G_1(s)}{\phi _2}\left( s \right)}}{{\left( {1 + {G_2}\left( s \right)} \right)\left( {1 + {L_1}\left( s\right)} \right)}} ,
\end{align}
where
\begin{align}\label{TrnasferOL1}
{L_1}\left( s  \right) = [1 - \frac{{{G_2}(s)}}{{1 + {G_2}(s )}}{e^{ - s \tau }}]{G_1}(s) = T_1(s ){G_1}(s ) 
\end{align}
gives the open-loop transfer function for the master laser.

In the above discussions, we have primarily focused on the laser phase noise $\phi$. 
It is readily verified that at the limit $G_1\gg 1$ and $G_2 \gg 1$, the laser phase noise is efficiently reduced by a factor of $L_1$.
However, a series of zeros inhabited in the transfer function Eq.~\eqref{TrnasferF1} and subsequently Eq.~\eqref{TrnasferOL1}, at $f  = n/ {\tau}$ where $n$ is an integer, will substantially undermines the efficiency of the suppression scheme.
To solve this issue, dual-arm locking schemes are proposed~\cite{arm-dual-2008}.
In the case of dual-arm locking, the arm-locking architecture is redesigned by using an appropriate combination of the interferometric data involving two arms.
As a result, the first zero corresponds to the difference between the two arm lengths, which can be pushed out of the frequency domain of observational interest.

\subsection{A real-time compensation scheme for the optical bench motion noise using the acousto-optic modulation}\label{section4.2}

\begin{figure}[!t]
	\includegraphics[width=0.50\textwidth]{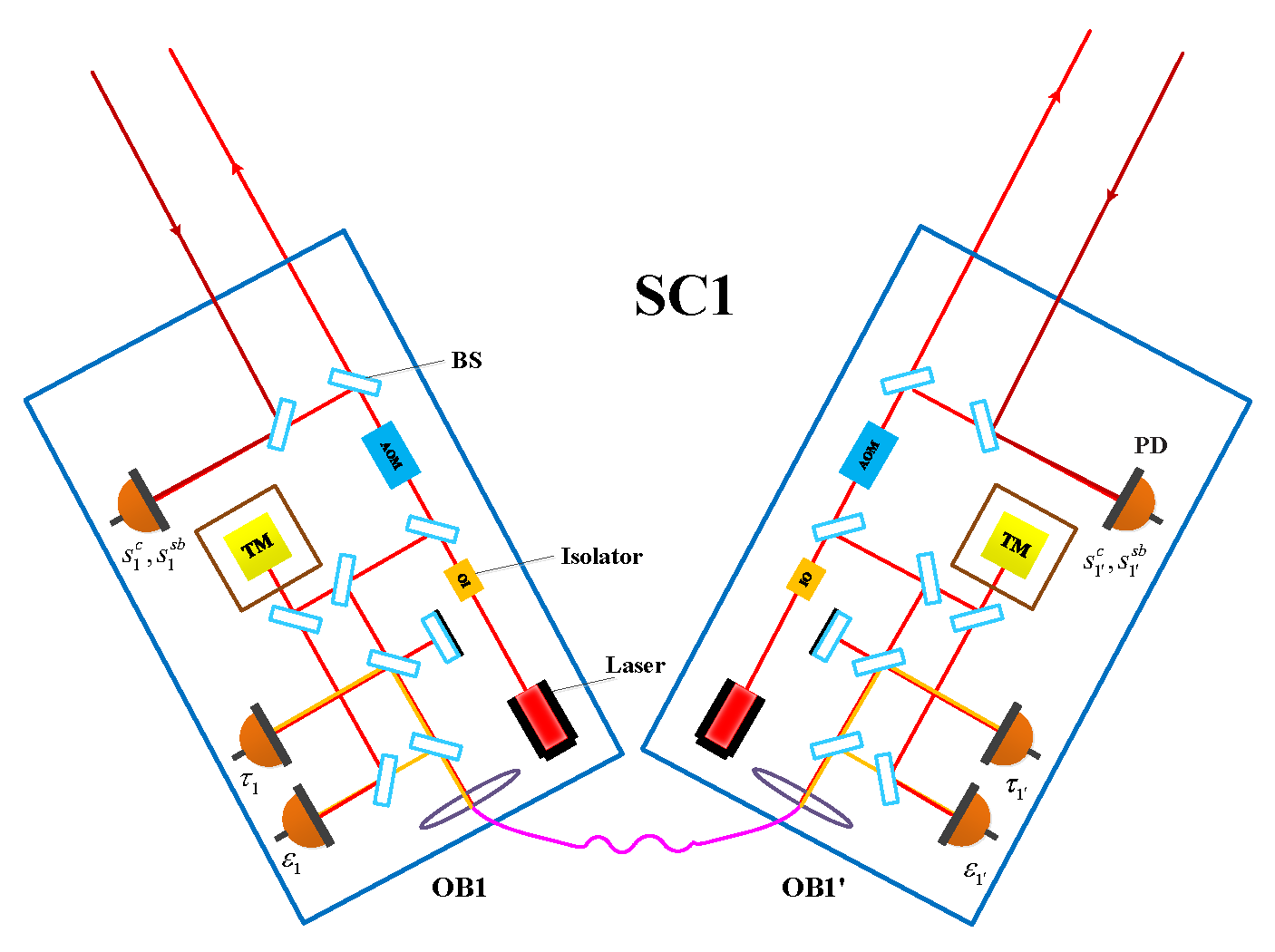}
	\caption{\label{fig5}
	The Schematic optical routing in the three spacecraft with real-time compensation of the optical bench vibration noise. SC: spacecraft; OB: optical bench; BS: beam splitter; PD: photodetector; IO: isolator; AOM: acousto-optic modulator; TM: test mass.}
\end{figure}

As discussed in Sec.~\ref{section3}, the TDI combinations are established in terms of the data streams given by in Eqs.~\eqref{postpre} and~\eqref{postpre1}.
In this subsection, we elaborate on a compensation scheme by manipulating the above data streams, which can be implemented with the arm locking technique to suppress the optical bench motion noise.

It is plausible to assume that the fiber noise is reciprocal~\cite{tdi-otto-2015}, namely, $\mu_{i'}=\mu_{i}$, 
As a result, it is not necessary to distinguish $\phi_{i}$ from $\phi_{i'}$~\cite{tdi-03}.
In practice, this can be effectively implemented by appropriately synchronizing the two lasers installed at the local spacecraft.

To proceed, one subtracts Eqs.~\eqref{t1} and~\eqref{tp1} respectively from Eqs.~\eqref{v1} and ~\eqref{vp1}, which gives rise to the following intermediate variables
\begin{subequations}
	\begin{align}
		a_i=&\frac{{{\varepsilon _i} - {r _i}}}{2} = - 2\pi {\nu _{i'}}\left[ {{\vec n_{(i - 1)'}} \cdot {\vec \delta _i} - {\vec n_{(i - 1)'}} \cdot {\vec \Delta _i}} \right]\notag\\
		 \equiv&{\Delta _i} - {\delta _i},\label{vartau1}\\
		a_{i'}=&\frac{\varepsilon _{i'} - r _{i'}}{2} =  - 2\pi {\nu _i}\left[ {{\vec n_{i + 1}} \cdot {\vec \delta _{i'}} - {\vec n_{i + 1}} \cdot {\vec \Delta _{i'}}} \right]\notag\\
		  \equiv& {\Delta _{i'}} - {\delta _{i'}}.\label{vartau1p}
	\end{align}
\end{subequations}
where one has introduced the shorthands
\begin{align}\label{hjdelta}
\delta _i=&2\pi\nu _{i'}\vec n_{(i - 1)'} \cdot {\vec \delta _i},\notag\\
\Delta _i=&2\pi\nu _{i'}\vec n_{(i - 1)'} \cdot {\vec \Delta _i},\notag\\
\delta _{i'}=&2\pi \nu _i \vec n_{i + 1} \cdot \vec \delta _{i'},\notag\\
\Delta _{i'}=&2\pi \nu _i \vec n_{i + 1} \cdot \vec \Delta _{i'}.
\end{align}
We propose the following compensation scheme based on the hard-wired AOM to compensate for the optical bench motion noise in the laser beams, as shown in Fig.~\ref{fig5}. 
In this scheme, two AOMs are installed to introduce a negative contribution governed by Eqs.~\eqref{vartau1} and~\eqref{vartau1p} to the science data stream.
Mathematically, one effectively modifies the definitions of the TDI variables $\eta_i,\eta_{i'}$ as follows
\begin{subequations}
	\begin{align}
		{\eta _i} =& s_i^c - a_i - {\cal D}_{i - 1} a_{(i+1)'}\notag\\
		=& s_i^c - \frac{{\varepsilon _i} - {r _i}}{2} - {{\cal D}_{i - 1}}\frac{{{\varepsilon _{(i + 1)'}} - {r _{(i + 1)'}}}}{2} \notag\\
		=& {{\cal D}_{i - 1}}\left[ {{\phi_{i + 1}} + {\delta _{(i + 1)'}}} \right] - \left[ {{\phi_i} + {\delta _i}} \right] + 2{\delta _i} + N_i^{S} + {h_i},\label{etai}\\
		{\eta _{i'}} =& s_{i'}^c - a_{i'} - {\cal D}_{(i+1)'} a_{i-1}\notag\\
		=& s_{i'}^c - \frac{{\varepsilon _{i'}} - {r _{i'}}}{2} - {{\cal D}_{(i + 1)'}}\frac{{{\varepsilon _{i - 1}} - {r_{i - 1}}}}{2} \notag\\
		=& {{\cal D}_{(i + 1)'}}\left[ {{\phi_{i - 1}} + {\delta _{i - 1}}} \right] - \left[ {{\phi_i} + {\delta _{i'}}} \right] + 2{\delta _{i'}} + N_{i'}^{S} + {h_{i'}}.\label{etaip}
	\end{align}
\end{subequations}

One observes that the spacecraft optical bench phase noise $\Delta _i$ has been eliminated from the above expressions.
Indeed, AOMs have been widely used in the feed-forward control of the laser frequency~\cite{chen2017-AOM}.
One also notes that the mathematical outcome of such cancelation is mainly reminiscent of the process introduced in Sec.~\ref{section3} regarding the TDI algorithm.
Therefore, such a hard-wired scheme can be used to replace the steps introduced in Eqs.~\eqref{postpre} and~\eqref{postpre1} for the post-processing stage.
Subsequently, the resulting noise floor of the arm locking will be determined mainly by the test mass noise and the shot noise, while the optical bench vibration noise is expected to be significantly suppressed by the procedure introduced here.

To further adopt a somewhat simplied notations and in particular, in accordance with the sign convention used in the arm locking formulation discussed above in Sec.~\ref{section4.1}, one further defines
\begin{align}\label{new}
    \tilde \eta_{i} \equiv& -\eta_{i},\notag\\
	\tilde \phi_{i} \equiv &{\phi_i} + {\delta _i},\notag\\
	{{\tilde \delta }_{i}} \equiv &{\delta_{i'}} - {\delta _{i}},\notag\\
	\tilde h_i\equiv & 2{\delta _i} + N_i^{S} + {h_i},\notag\\
	\tilde h_{i'}\equiv & 2{\delta _{i'}} + N_{i'}^{S} + {h_{i'}},
\end{align}
where
\begin{align}
	\tilde \phi_{i'} \equiv &{\phi_{i'}} + {\delta _{i'}} =\tilde\phi_i + \tilde\delta_i \notag
\end{align}
is implied.

\begin{figure}[!t]
\includegraphics[width=0.50\textwidth]{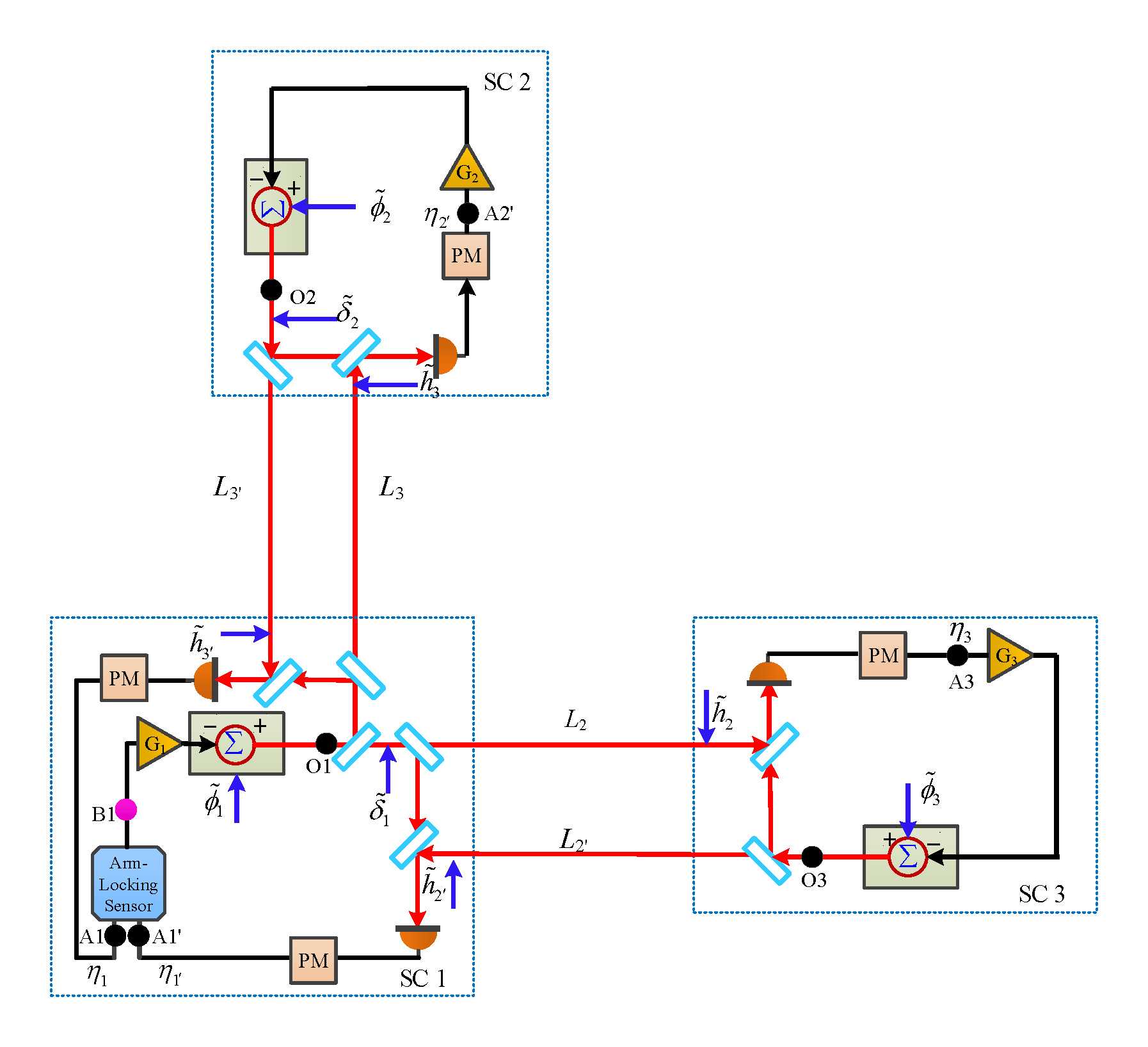}
\caption{\label{fig6} 
Schematic optical routing for dual-arm locking. SC: spacecraft; OB: optical bench; BS: beam splitter; PM: phasemeter; PD: photodetector; IO: isolator; AOM: acousto-optic modulator; TM: test mass.}
\end{figure}

By using Eq.~\eqref{new}, the TDI observables Eqs.~\eqref{etai} and~\eqref{etaip} can be rewritten as
\begin{subequations}
\begin{align}
	\tilde\eta _i =&  {{\tilde \phi}_{i}} - {{\cal D}_{i - 1}}\left( \tilde \phi_{(i + 1)} + \tilde \delta _{i + 1} \right) - \tilde h_i,\label{newetai}\\
	\tilde\eta _{i'} =& \left( \tilde \phi_{i} + \tilde \delta_i \right) - {{\cal D}_{(i + 1)'}}{{\tilde \phi}_{(i - 1)}} - \tilde h_{i'} ,\label{newetaipie}
\end{align}
\end{subequations}
which may siginificant simplfy the symbolic noations used in, for instance, Fig.~\ref{fig6}, and therefore will be adopted in the remainder of the paper.
Despite its simplicity, We note that the apparent symmetry between laser $i$ and $i'$ is lost in the above notation.

\subsection{Output observables from arm locking signal routing}\label{section4.3}

We can now derive the forms of data streams by adopting the above compensation scheme and the dual-arm locking technique.
One considers spacecraft 1 (denoted by SC1) shown in Fig.~\ref{fig6} as the master spacecraft.
Again, we start with the simplest scenario.
If the phase locking loops at the spacecraft 1 and 3 are both open and one only considers the laser phase noise, the interferometric output at A3 is given by
\begin{align}\label{ETA3}
\tilde\eta_3 = {\phi _3}(s) - {\phi _1}(s){e^{ - s{\tau _2}}}.
\end{align}
In the framework of the first generation TDI, one denotes that $\frac{\tau _3 + \tau _2}{2} = \frac{\tau }{2},\frac{\tau _3 - \tau _2}{2} = \frac{\Delta \tau }{2}$,
and approximates the time delay along the three arms as
\begin{align}\label{tao123}
{\tau _1} = &\tau _{1'} = \frac{\tau  + \delta \tau }{2},\notag\\
{\tau _2} =& \tau _{2'} = \frac{\tau  - \Delta \tau }{2},\notag\\
{\tau _3} =& \tau _{3'} = \frac{\tau  + \Delta \tau }{2},
\end{align}
where one bears in mind that $\Delta \tau  \ll \tau ,\delta \tau  \ll \tau$.

In addition, one needs to consider both the arm locking discussed in Sec.~\ref{section4.1} and the compensation scheme of Sec.~\ref{section4.2}. 

If the laser at spacecraft 3 is phase locked to the incoming laser, the signal at $O3$ is given by
\begin{align}\label{phio3}
{\phi _{O3}} = \frac{{{\tilde \phi _{3}}}}{{1 + {G_3}}} + \frac{{{G_3}}}{{1 + {G_3}}}\left[ {\left( {{\phi _{O1}} + {{\tilde \delta }_1}} \right){e^{ - \frac{s}{2}(\tau  - \Delta \tau )}} + {\tilde h_3}} \right],
\end{align}
Similarly, one can writes down the remaining terms.
For instance, the signal at $O2'$ reads
\begin{align}\label{phio2}
{\phi_{O2'}}\equiv{\phi_{O2}} + {\tilde \delta _2} = \frac{\tilde \phi _{2} + \tilde \delta_2}{1 + G_2} + \frac{{{G_2}}}{{1 + {G_2}}}\left[ {{\phi _{O1}}{e^{ - \frac{s}{2}(\tau  + \Delta \tau )}} + {\tilde h_{2'}}} \right],
\end{align}
where $G_i$ is the controller gain of the phase-locking loop at spacecraft $i$.

Furthermore, if the phase locking loop is open at master spacecraft (SC1), i.e., $G_1=0$, the signal at $A_1$ and $A_{1'}$ can be expressed as, in accordance with Eqs.~\eqref{newetai} and~\eqref{newetaipie},
\begin{align}\label{armeta1}
\tilde\eta_1 =& {\phi _{O1}} - \left[ {\left( {{\phi _{O2}} + {{\tilde \delta }_2}} \right){e^{ - \frac{s}{2}(\tau  + \Delta \tau )}} + {\tilde h_1}} \right]\notag\\
 =& {\phi _{O1}}{T_3} - \frac{{{\tilde \phi _{2}} + {{\tilde \delta }_2}}}{{1 + {G_2}}}{e^{ - \frac{s}{2}(\tau  + \Delta \tau )}}\notag\\
  -& \left[ {\frac{{{G_2}}}{{1 + {G_2}}}{\tilde h_{2'}}{e^{ - \frac{s}{2}(\tau  + \Delta \tau )}} + {\tilde h_1}} \right],
  \end{align}
  and
  \begin{align}\label{armeta1pie}
\tilde\eta_{1'} =& \phi_{O1'} - \left[ {{\phi _{O3}}{e^{ - \frac{s}{2}(\tau  - \Delta \tau )}} + {h_{{\rm{1'}}}}} \right]\notag\\
 =& \left( {{\phi _{O1}} + {{\tilde \delta }_1}} \right) - \left[ {{\phi _{O3}}{e^{ - \frac{s}{2}(\tau  - \Delta \tau )}} + {h_{{\rm{1'}}}}} \right]\notag\\
 =& \left( {{\phi _{O1}} + {{\tilde \delta }_1}} \right){T_{\rm{2}}} - \frac{{{\tilde \phi _{3}}}}{{1 + {G_3}}}{e^{ - \frac{s}{2}(\tau  - \Delta \tau )}}\notag\\
 -& \left[ {\frac{{{G_3}}}{{1 + {G_3}}}{\tilde h_{\rm{3}}}{e^{ - \frac{s}{2}(\tau  - \Delta \tau )}} + {\tilde h_{{\rm{1'}}}}} \right],
\end{align}
where $T_2$ and $T_3$ are the frequency response corresponding to arm-length $L_2$ and $L_3$.
\begin{align}\label{t2t3}
{T_3} =& 1 - \frac{{{G_2}}}{{1 + {G_2}}}{e^{ - s(\tau  + \Delta \tau )}},\notag\\
{T_{2}} =& 1 - \frac{{{G_3}}}{{1 + {G_3}}}{e^{ - s(\tau  - \Delta \tau )}}.
\end{align}

It is convenient to rewrite the above results regarding $\tilde\eta_{1}$ and $\tilde\eta_{1'}$ in terms of a two-component column matrix
\begin{align}\label{etaA1ding}
\tilde \eta _{A1}= \left[ {\begin{array}{*{20}{c}}
{{\tilde \eta _1}}\\
{{\tilde \eta _{1'}}}
\end{array}} \right] .
\end{align}
To be specific
\begin{align}\label{etaA1}
{\tilde \eta _{A1}} =& {\phi _{O1}}\left[{\begin{array}{*{20}{c}}
{T_3}\\
{T_2}
\end{array}} \right]- \left[{\begin{array}{*{20}{c}}
{\frac{{{\tilde \phi _{2}} + {{\tilde \delta }_2}}}{{1 + {G_2}}}{e^{ - \frac{s}{2}(\tau  + \Delta \tau )}}}\\
{\frac{{{\tilde \phi _{3}}}}{{1 + {G_3}}}{e^{ - \frac{s}{2}(\tau  - \Delta \tau )}}}
\end{array}} \right] \notag\\
-& \left[ {\begin{array}{*{20}{c}}
{\frac{{{G_2}}}{{1 + {G_2}}}{\tilde h_{{\rm{2'}}}}{e^{ - \frac{s}{2}(\tau  + \Delta \tau )}} + {\tilde h_{\rm{1}}}}\\
{\frac{{{G_3}}}{{1 + {G_3}}}{\tilde h_{\rm{3}}}{e^{ - \frac{s}{2}(\tau  - \Delta \tau )}} + {\tilde h_{{\rm{1'}}}}}
\end{array}} \right] + \left[{\begin{array}{*{20}{c}}
0\\
\tilde \delta _1 T_2
\end{array}} \right].
\end{align}

Now, we move on to deal with specific arm locking implementations.
Following~\cite{arm-moddual-2009}, one introduces a mapping function $S_X$ to distinguish the different types of arm locking.
It is a two-component row matrix which essentially performs an appropriate combination of the two components of $\tilde \eta_{A1}$, so that the output at the point $B_1$ is formally expressed as
\begin{align}\label{phiB1}
{\left. {{\phi _{B1}}} \right|_X} = {{\rm{S}}_X}{\tilde \eta _{A1}}.
\end{align}

In the case of dual-arm locking, the mapping function reads
\begin{align}\label{SD}
{S_D} = \left[ {\begin{array}{*{20}{c}}
{1{\rm{ + }}\frac{1}{{s\Delta \tau }},}&{1 - \frac{1}{{s\Delta \tau }}}
\end{array}} \right].
\end{align}
The output at $O1$ reads
\begin{align}\label{SDO1}
{\phi _{O1}} = {\tilde \phi _{1}} - {G_1}{S_D}{\tilde \eta _{A1}} .
\end{align}
By substituting Eq.~\eqref{etaA1} into Eq.~\eqref{SDO1} and solving for $\phi_{O1}$, we have
\begin{widetext}
\begin{align}\label{phio1}
{\phi _{O1}} =& \frac{{{\tilde \phi _{1}}}}{{1 + {G_1}{T_D}}} + \left[ {\frac{{({\tilde \phi _{2}} + {{\tilde \delta }_2}){e^{ - \frac{s}{2}(\tau  + \Delta \tau )}}}}{{1 + {G_2}}} + \frac{{{G_2}}}{{1 + {G_2}}}{\tilde h_{2'}}{e^{ - \frac{s}{2}(\tau  + \Delta \tau )}} + {\tilde h_1}} \right]\frac{{{G_1}}}{{1 + {G_1}{T_D}}}\left( {1 + \frac{1}{{s\Delta \tau }}} \right)\notag\\
+& \left[ {\frac{{{\tilde \phi _{3}}{e^{ - \frac{s}{2}(\tau  - \Delta \tau )}}}}{{1 + {G_3}}} + \frac{{{G_3}}}{{1 + {G_3}}}{\tilde h_3}{e^{ - \frac{s}{2}(\tau  - \Delta \tau )}} + {\tilde h_{1'}} - {{\tilde \delta }_1}{T_2}} \right]\frac{{{G_1}}}{{1 + {G_1}{T_D}}}\left( {1 - \frac{1}{{s\Delta \tau }}} \right),
\end{align}
where
\begin{align}\label{td}
{T_D} = {T_3} + {T_2} + \frac{1}{{s\Delta \tau }}({T_3} - {T_2}) = {T_ + } + \frac{1}{{s\Delta \tau }}{T_ - }.
\end{align}
By substituting Eq.~\eqref{phio1} back into Eq.~\eqref{etaA1}, one finds
\begin{align}\label{PHIA1o}
{\tilde \eta _{A1}} =& \frac{{{\tilde \phi _{1}}}}{{1 + {G_1}{T_D}}}\left[ {\begin{array}{*{20}{c}}
{{T_3}}\\
{{T_2}}
\end{array}} \right] + \frac{{({\tilde \phi _{2}} + {{\tilde \delta }_2}){e^{ - \frac{s}{2}(\tau  + \Delta \tau )}}}}{{1 + {G_2}}}\left[ {\begin{array}{*{20}{c}}
{\frac{{{G_1}}}{{1 + {G_1}{T_D}}}\left( {1 + \frac{1}{{s\Delta \tau }}} \right){T_3} - 1}\\
{\frac{{{G_1}}}{{1 + {G_1}{T_D}}}\left( {1 + \frac{1}{{s\Delta \tau }}} \right){T_2}}
\end{array}} \right]\notag\\
 +& \frac{{{\tilde \phi _{3}}{e^{ - \frac{s}{2}(\tau  - \Delta \tau )}}}}{{1 + {G_3}}}\left[ {\begin{array}{*{20}{c}}
{\frac{{{G_1}}}{{1 + {G_1}{T_D}}}\left( {1 - \frac{1}{{s\Delta \tau }}} \right){T_3}}\\
{\frac{{{G_1}}}{{1 + {G_1}{T_D}}}\left( {1 - \frac{1}{{s\Delta \tau }}} \right){T_2} - 1}
\end{array}} \right]\notag\\
 +& \left( {\frac{{{G_2}}}{{1 + {G_2}}}{\tilde h_{2'}}{e^{ - \frac{s}{2}(\tau  + \Delta \tau )}} + {\tilde h_1}} \right)\left[ {\begin{array}{*{20}{c}}
{\frac{{{G_1}}}{{1 + {G_1}{T_D}}}\left( {1 + \frac{1}{{s\Delta \tau }}} \right){T_3} - 1}\\
{\frac{{{G_1}}}{{1 + {G_1}{T_D}}}\left( {1 + \frac{1}{{s\Delta \tau }}} \right){T_2}}
\end{array}} \right]\notag\\
 +& \left( {\frac{{{G_3}}}{{1 + {G_3}}}{\tilde h_3}{e^{ - \frac{s}{2}(\tau  - \Delta \tau )}} + {\tilde h_{1'}} - {{\tilde \delta }_1}{T_2}} \right)\left[ {\begin{array}{*{20}{c}}
{\frac{{{G_1}}}{{1 + {G_1}{T_D}}}\left( {1 - \frac{1}{{s\Delta \tau }}} \right){T_3}}\\
{\frac{{{G_1}}}{{1 + {G_1}{T_D}}}\left( {1 - \frac{1}{{s\Delta \tau }}} \right){T_2} - 1}
\end{array}} \right],
\end{align}
where
\begin{align}
\frac{{{G_1}}}{{1 + {G_1}{T_D}}}\left( {1 + \frac{1}{{s\Delta \tau }}} \right){T_3} - 1 =&  - \frac{{{G_1}}}{{1 + {G_1}{T_D}}}\left[ {\left( {1 - \frac{1}{{s\Delta \tau }}} \right){T_2} + \frac{1}{{{G_1}}}} \right],\notag\\
\frac{{{G_1}}}{{1 + {G_1}{T_D}}}\left( {1 - \frac{1}{{s\Delta \tau }}} \right){T_2} - 1 =&  - \frac{{{G_1}}}{{1 + {G_1}{T_D}}}\left[ {\left( {1 + \frac{1}{{s\Delta \tau }}} \right){T_3} + \frac{1}{{{G_1}}}} \right].
\end{align}

By rewriting Eq.~\eqref{PHIA1o} in terms of its components, we obtained the desired results,
\begin{align}\label{PHI1}
\tilde\eta_1 =& \frac{{{{\tilde \phi }_1}}}{{1 + {G_1}{T_D}}}{T_3} + \left[ {\frac{{({{\tilde \phi }_2} + {{\tilde \delta }_2}){e^{ - \frac{s}{2}(\tau  + \Delta \tau )}}}}{{1 + {G_2}}} + \frac{{{G_2}}}{{1 + {G_2}}}{h_{2'}}{e^{ - \frac{s}{2}(\tau  + \Delta \tau )}} + {h_1}} \right]\frac{{{G_1}}}{{1 + {G_1}{T_D}}}\left[ { - \left( {1 - \frac{1}{{s\Delta \tau }}} \right){T_2} - \frac{1}{{{G_1}}}} \right]\notag\\
 +& \left[ {\frac{{{{\tilde \phi }_3}{e^{ - \frac{s}{2}(\tau  - \Delta \tau )}}}}{{1 + {G_3}}} + \frac{{{G_3}}}{{1 + {G_3}}}{h_3}{e^{ - \frac{s}{2}(\tau  - \Delta \tau )}} + {h_{1'}} - {{\tilde \delta }_1}{T_2}} \right]\frac{{{G_1}}}{{1 + {G_1}{T_D}}}\left( {1 - \frac{1}{{s\Delta \tau }}} \right){T_3},
\end{align}
and
\begin{align}\label{PHI1pie}
\tilde\eta_{1'} =& \frac{{{{\tilde \phi }_1}}}{{1 + {G_1}{T_D}}}{T_2} + \left[ {\frac{{({{\tilde \phi }_2} + {{\tilde \delta }_2}){e^{ - \frac{s}{2}(\tau  + \Delta \tau )}}}}{{1 + {G_2}}} + \frac{{{G_2}}}{{1 + {G_2}}}{h_{2'}}{e^{ - \frac{s}{2}(\tau  + \Delta \tau )}} + {h_1}} \right]\frac{{{G_1}}}{{1 + {G_1}{T_D}}}\left( {1 + \frac{1}{{s\Delta \tau }}} \right){T_2}\notag\\
 +& \left[ {\frac{{{{\tilde \phi }_3}{e^{ - \frac{s}{2}(\tau  - \Delta \tau )}}}}{{1 + {G_3}}} + \frac{{{G_3}}}{{1 + {G_3}}}{h_3}{e^{ - \frac{s}{2}(\tau  - \Delta \tau )}} + {h_{1'}} - {{\tilde \delta }_1}{T_2}} \right]\frac{{{G_1}}}{{1 + {G_1}{T_D}}}\left[ { - \left( {1 + \frac{1}{{s\Delta \tau }}} \right){T_3} - \frac{1}{{{G_1}}}} \right].
\end{align}

The calculations for the remaining variables can be carried out similarly in a straightforward but somewhat tedious fashion.
The resultant signals at point $A3$ and point $A2'$,
\begin{align}\label{eta333}
\tilde \eta _3 =& {\phi _{O3}} - \left[ {\left( {{\phi _{O1}} + {{\tilde \delta }_1}} \right){e^{ - \frac{s}{2}(\tau  - \Delta \tau )}} + {\tilde h_3}} \right] \notag\\
=& \frac{{{\tilde \phi _{3}}}}{{1 + {G_3}}} - \frac{{\left( {{\phi _{O1}} + {{\tilde \delta }_1}} \right){e^{ - \frac{s}{2}(\tau  - \Delta \tau )}}}}{{1 + {G_3}}} - \frac{{{\tilde h_3}}}{{1 + {G_3}}},\notag\\
\tilde \eta _{2'} =& \phi _{O2}+\tilde \delta_2- \left[ \phi _{O1}{e^{ - \frac{s}{2}(\tau + \Delta \tau )}} + \tilde h_{2'} \right] \notag\\
=& \frac{\tilde \phi _{2}+\tilde \delta _2}{1 + G_2} - \frac{\phi _{O1}e^{ - \frac{s}{2}(\tau + \Delta \tau)}}{1 + G_2} - \frac{\tilde h_{2'}}{1 + G_2} ,
\end{align}
are found to be
\begin{align}\label{eta3gg}
{\tilde \eta _3} =& \frac{{{{\tilde \phi }_3}}}{{(1 + {G_3})}} - \frac{{{h_3}}}{{(1 + {G_3})}} - \frac{{{{\tilde \delta }_1}{e^{ - \frac{s}{2}(\tau  - \Delta \tau )}}}}{{(1 + {G_3})}} - \frac{{{{\tilde \phi }_1}}}{{\left( {1 + {G_1}{T_D}} \right)(1 + {G_3})}}{e^{ - \frac{s}{2}(\tau  - \Delta \tau )}}\notag\\
 -& \frac{{{G_1}}}{{\left( {1 + {G_1}{T_D}} \right)(1 + {G_3})}}\left[ {\frac{{({{\tilde \phi }_2} + {{\tilde \delta }_2}){e^{ - \frac{s}{2}(\tau  + \Delta \tau )}}}}{{1 + {G_2}}}\left( {1 + \frac{1}{{s\Delta \tau }}} \right) + \frac{{{{\tilde \phi }_3}{e^{ - \frac{s}{2}(\tau  - \Delta \tau )}}}}{{1 + {G_3}}}\left( {1 - \frac{1}{{s\Delta \tau }}} \right)} \right]{e^{ - \frac{s}{2}(\tau  - \Delta \tau )}}\notag\\
 -& \frac{{{G_1}}}{{\left( {1 + {G_1}{T_D}} \right)(1 + {G_3})}}\left[ {\left( {\frac{{{G_2}}}{{1 + {G_2}}}{h_{2'}}{e^{ - \frac{s}{2}(\tau  + \Delta \tau )}} + {h_1}} \right)\left( {1 + \frac{1}{{s\Delta \tau }}} \right) + \left( {\frac{{{G_3}}}{{1 + {G_3}}}{h_3}{e^{ - \frac{s}{2}(\tau  - \Delta \tau )}} + {h_{1'}} - {{\tilde \delta }_1}{T_2}} \right)\left( {1 - \frac{1}{{s\Delta \tau }}} \right)} \right]{e^{ - \frac{s}{2}(\tau  - \Delta \tau )}},
\end{align}
and
\begin{align}\label{eta2pgg}
{\tilde \eta _{2'}} =& \frac{{{{\tilde \phi }_2} + {{\tilde \delta }_2}}}{{(1 + {G_2})}} - \frac{{{h_{2'}}}}{{(1 + {G_2})}} - \frac{{{{\tilde \phi }_1}}}{{\left( {1 + {G_1}{T_D}} \right)(1 + {G_2})}}{e^{ - \frac{s}{2}(\tau  + \Delta \tau )}}\notag\\
 -& \frac{{{G_1}}}{{\left( {1 + {G_1}{T_D}} \right)(1 + {G_2})}}\left[ {\frac{{({{\tilde \phi }_2} + {{\tilde \delta }_2}){e^{ - \frac{s}{2}(\tau  + \Delta \tau )}}}}{{1 + {G_2}}}\left( {1 + \frac{1}{{s\Delta \tau }}} \right) + \frac{{{{\tilde \phi }_3}{e^{ - \frac{s}{2}(\tau  - \Delta \tau )}}}}{{1 + {G_3}}}\left( {1 - \frac{1}{{s\Delta \tau }}} \right)} \right]{e^{ - \frac{s}{2}(\tau  + \Delta \tau )}}\notag\\
 -& \frac{{{G_1}}}{{\left( {1 + {G_1}{T_D}} \right)(1 + {G_2})}}\left[ {\left( {\frac{{{G_2}}}{{1 + {G_2}}}{h_{2'}}{e^{ - \frac{s}{2}(\tau  + \Delta \tau )}} + {h_1}} \right)\left( {1 + \frac{1}{{s\Delta \tau }}} \right) + \left( {\frac{{{G_3}}}{{1 + {G_3}}}{h_3}{e^{ - \frac{s}{2}(\tau  - \Delta \tau )}} + {h_{1'}} - {{\tilde \delta }_1}{T_2}} \right)\left( {1 - \frac{1}{{s\Delta \tau }}} \right)} \right]{e^{ - \frac{s}{2}(\tau  + \Delta \tau )}}.
 \end{align}

Lastly, the output at point $A3'$ and point $A2$,
\begin{align}\label{}
{\tilde \eta _{3'}}  =& {\phi _{O3}} + {{\tilde \delta }_3} - \left( {{\phi _{O2}}{e^{ - \frac{s}{2}(\tau  + \delta \tau )}} + {\tilde h_{3'}}} \right),\notag\\
{\tilde \eta _2} =& {\phi _{O2}} - \left[ {({\phi _{O3}} + {{\tilde \delta }_3}){e^{ - \frac{s}{2}(\tau  + \delta \tau )}} + {\tilde h_2}} \right].
\end{align}
can be expressed as
\begin{align}\label{eta3pie}
{\tilde \eta _{3'}} =&\frac{{{\tilde \phi _{1}}}}{{1 + {G_1}{T_D}}}\left({\frac{{{G_3}{e^{ - \frac{s}{2}(\tau  - \Delta \tau )}}}}{{1 + {G_3}}} - \frac{{{G_2}{e^{ - \frac{s}{2}(2\tau  + \Delta \tau  + \delta \tau )}}}}{{1 + {G_2}}}} \right)\notag\\
 +& \frac{{({\tilde \phi _{2}} + {{\tilde \delta }_2}){e^{ - \frac{s}{2}(\tau  + \delta \tau )}}}}{{1 + {G_2}}}\left[ { - 1 + \frac{{{G_1}}}{{1 + {G_1}{T_D}}}\left( {\frac{{{G_3}{e^{ - \frac{s}{2}(\tau  - \delta \tau )}}}}{{1 + {G_3}}} - \frac{{{G_2}{e^{ - s(\tau  + \Delta \tau )}}}}{{1 + {G_2}}}} \right)\left( {1 + \frac{1}{{s\Delta \tau }}} \right)} \right]\notag\\
 +&\frac{{{\tilde \phi _{3}}}}{{1 + {G_3}}}\left[ {1 + \frac{{{G_1}}}{{1 + {G_1}{T_D}}}\left( {\frac{{{G_3}{e^{ - s(\tau  - \Delta \tau )}}}}{{1 + {G_3}}} - \frac{{{G_2}{e^{ - \frac{s}{2}(3\tau  + \delta \tau )}}}}{{1 + {G_2}}}} \right)\left( {1 - \frac{1}{{s\Delta \tau }}} \right)} \right]\notag\\
 +& \frac{{{G_1}}}{{1 + {G_1}{T_D}}}\left( {\frac{{{G_3}{e^{ - \frac{s}{2}(\tau  - \Delta \tau )}}}}{{1 + {G_3}}} - \frac{{{G_2}{e^{ - \frac{s}{2}(2\tau  + \Delta \tau  + \delta \tau )}}}}{{1 + {G_2}}}} \right)\left[ \begin{array}{l}
\left( {1 + \frac{1}{{s\Delta \tau }}} \right)\left( {\frac{{{G_2}}}{{1 + {G_2}}}{\tilde h_{2'}}{e^{ - \frac{s}{2}(\tau  + \Delta \tau )}} + {\tilde h_1}} \right)\\
 + \left( {1 - \frac{1}{{s\Delta \tau }}} \right)\left( {\frac{{{G_3}}}{{1 + {G_3}}}{\tilde h_3}{e^{ - \frac{s}{2}(\tau  - \Delta \tau )}} + {\tilde h_{1'}} - {{\tilde \delta }_1}{T_2}} \right)
\end{array} \right]\notag\\
 +& \frac{{{G_3}}}{{1 + {G_3}}}{\tilde h_3} - \frac{{{G_2}}}{{1 + {G_2}}}{\tilde h_{2'}}{e^{ - \frac{s}{2}(\tau  + \delta \tau )}} - {\tilde h_{3'}} + \frac{{{G_3}}}{{1 + {G_3}}}{{\tilde \delta }_1}{e^{ - \frac{s}{2}(\tau  - \Delta \tau )}} + {{\tilde \delta }_2}{e^{ - \frac{s}{2}(\tau  + \delta \tau )}} + {{\tilde \delta }_3},
\end{align}
and
\begin{align}\label{eta22222}
{\tilde \eta _2} =& \frac{{{\tilde \phi _{1}}}}{{1 + {G_1}{T_D}}}\left( {\frac{{{G_2}{e^{ - \frac{s}{2}(\tau  + \Delta \tau )}}}}{{1 + {G_2}}} - \frac{{{G_3}{e^{ - \frac{s}{2}(2\tau  - \Delta \tau  + \delta \tau )}}}}{{1 + {G_3}}}} \right)\notag\\
+&\frac{{{\tilde \phi _{2}} + {{\tilde \delta }_2}}}{{1 + {G_2}}}\left[ {1 + \frac{{{G_1}}}{{1 + {G_1}{T_D}}}\left( {\frac{{{G_2}{e^{ - s(\tau  + \Delta \tau )}}}}{{1 + {G_2}}} - \frac{{{G_3}{e^{ - \frac{s}{2}(3\tau  + \delta \tau )}}}}{{1 + {G_3}}}} \right)\left( {1 + \frac{1}{{s\Delta \tau }}} \right)} \right] \notag\\
 +& \frac{{{\tilde \phi _{3}}}}{{1 + {G_3}}}{e^{ - \frac{s}{2}(\tau  + \delta \tau )}}\left[ { - 1 + \frac{{{G_1}}}{{1 + {G_1}{T_D}}}\left( {\frac{{{G_2}{e^{ - \frac{s}{2}(\tau  - \delta \tau )}}}}{{1 + {G_2}}} - \frac{{{G_3}{e^{ - \frac{s}{2}(2\tau  - 2\Delta \tau )}}}}{{1 + {G_3}}}} \right)\left( {1 - \frac{1}{{s\Delta \tau }}} \right)} \right]\notag\\
 +& \frac{{{G_1}}}{{1 + {G_1}{T_D}}}\left( {\frac{{{G_2}{e^{ - \frac{s}{2}(\tau  + \Delta \tau )}}}}{{1 + {G_2}}} - \frac{{{G_3}{e^{ - \frac{s}{2}(2\tau  - \Delta \tau  + \delta \tau )}}}}{{1 + {G_3}}}} \right)\left[ \begin{array}{l}
\left( {1 + \frac{1}{{s\Delta \tau }}} \right)\left( {\frac{{{G_2}}}{{1 + {G_2}}}{\tilde h_{2'}}{e^{ - \frac{s}{2}(\tau  + \Delta \tau )}} + {\tilde h_1}} \right)\\
 + \left( {1 - \frac{1}{{s\Delta \tau }}} \right)\left({\frac{{{G_3}}}{{1 + {G_3}}}{\tilde h_3}{e^{ - \frac{s}{2}(\tau  - \Delta \tau )}} + {\tilde h_{1'}} - {{\tilde \delta }_1}{T_{2}}} \right)
\end{array} \right]\notag\\
 + &\frac{{{G_2}}}{{1 + {G_2}}}{\tilde h_{2'}} - \frac{{{G_3}}}{{1 + {G_3}}}{\tilde h_3}{e^{ - \frac{s}{2}(\tau  + \delta \tau )}} - {\tilde h_2} - \frac{{{G_3}}}{{1 + {G_3}}}{{\tilde \delta }_1}{e^{ - \frac{s}{2}(2\tau  - \Delta \tau  + \delta \tau )}} - {{\tilde \delta }_3}{e^{ - \frac{s}{2}(\tau  + \delta \tau )}} - {{\tilde \delta }_2} .
\end{align}
where the suppression factors for the GW signals and noise are similar to the case of $\tilde\eta_1$ and $\tilde\eta_{1'}$.

We note that the obtained observables are distinct from those previously given by either Eqs.~\eqref{newetai} and~\eqref{newetaipie} or Eqs.~\eqref{postpre} and~\eqref{postpre1}.
The main difference is that the present results depend on the specific form of arm locking scheme.
For $\tilde\eta_1$, the magnitude of the laser phase noise $\phi _{1}$ is considerably suppressed by a factor of $1 + {G_1}{T_D}$.
In addition, the laser phase noise $\phi _{2},\phi _{3}$ also enter the variable $\tilde\eta_1$ while being suppressed by a factor of similar form. 
The GW signals and other types of noise remain their presence.
While the orders of magnitude of these quantities remain unchanged, their forms are distorted, governed by the specific arm locking scenario.
Besides, it is noted that fractions of GW signal and noise associated with other arm lengths and spacecraft have also sneaked into the expression. 
This feature is shared by the varialbles $\tilde\eta_{1'}$, $\tilde\eta_{2}$, and $\tilde\eta_{3'}$.
It is also worth noting that for the observables $\tilde\eta_{2'}$ and $\tilde\eta_{3}$, it is less desirable that both the noise terms and GW signals in Eqs.~\eqref{eta333} are significantly suppressed approximately by a factor $1+G_i$.
In other words, besides the laser phase noise, the GW signal is significantly suppressed in these variables due to the arm locking optical routing.

The obtained variables $\tilde\eta_i$ given by Eqs.~\eqref{PHI1},~\eqref{PHI1pie},~\eqref{eta3gg},~\eqref{eta2pgg},~\eqref{eta3pie}, and~\eqref{eta22222} furnish the observables that can be used to search for the relevant TDI combinations.
\end{widetext}

\section{TDI solutions for the arm-locking variables}\label{section5}

\subsection{Simplified forms of the TDI observables}\label{section5.1}

Before proceeding to construct the TDI solutions, we first simplify the expressions for the variables obtained in the last section.
Regarding the arm locking architecture design, the controller's gain is typically rather significant, i.e., $G_i\gg 1$.
On the other hand, $\Delta \tau $, which is governed by the difference between two armlengths, is relatively insignificant.
As a result, for moderate GW frequency, $\omega \Delta \tau  \ll 1$.
Therefore, it is plausible to make the following approximations
\begin{align}\label{tftt}
&{T_2} \approx T_3 = T = 1 - {e^{ - s\tau }},\notag\\
&{T_D} \approx 2.
\end{align}
Moreover, one may further ignore the differences between the time delays when they appear in the exponentials of the relevant terms, namely, 
\begin{align}\label{tauapp}
\tau_1 \approx \tau_2 \approx \tau_3 = \frac{\tau}{2} .
\end{align}
However, a time difference shown up elsewhere should be kept and evaluated according to Eq.~\eqref{tao123}.

By substituting Eq.~\eqref{tftt} into Eqs.~\eqref{PHI1},~\eqref{PHI1pie},~\eqref{eta3gg},~\eqref{eta2pgg},~\eqref{eta3pie}, and~\eqref{eta22222}, these expressions are significantly simplified to give
\begin{align}\label{finallyeta1}
{\tilde \eta _1} \approx &\frac{{{{\tilde \phi }_1}}}{{2{G_1}}}\left( {1 - {e^{ - s\tau }}} \right)\notag\\
 -& \frac{1}{2}\left( {\frac{{({{\tilde \phi }_2} + {{\tilde \delta }_2}){e^{ - \frac{s}{2}\tau }}}}{{{G_2}}} - \frac{{{{\tilde \phi }_3}{e^{ - \frac{s}{2}\tau }}}}{{{G_3}}}} \right)\left( {1 - \frac{1}{{s\Delta \tau }}} \right)\left( {1 - {e^{ - s\tau }}} \right)\notag\\
 +& \frac{1}{2}\left( { - {{\tilde h}_{2'1}} + {{\tilde h}_{31'}} - {{\tilde \delta }_1}\left( {1 - {e^{ - s\tau }}} \right)} \right)\left( {1 - \frac{1}{{s\Delta \tau }}} \right)\left( {1 - {e^{ - s\tau }}} \right),
\end{align}
\begin{align}\label{finallyeta1pie}
{\tilde \eta _{1'}} \approx& \frac{{{{\tilde \phi }_1}}}{{2{G_1}}}\left( {1 - {e^{ - s\tau }}} \right)\notag\\
 +& \frac{1}{2}\left( {\frac{{({{\tilde \phi }_2} + {{\tilde \delta }_2}){e^{ - \frac{s}{2}\tau }}}}{{{G_2}}} - \frac{{{{\tilde \phi }_3}{e^{ - \frac{s}{2}\tau }}}}{{{G_3}}}} \right)\left( {1 + \frac{1}{{s\Delta \tau }}} \right)\left( {1 - {e^{ - s\tau }}} \right)\notag\\
 -& \frac{1}{2}\left( { - {{\tilde h}_{2'1}} + {{\tilde h}_{31'}} - {{\tilde \delta }_1}\left( {1 - {e^{ - s\tau }}} \right)} \right)\left( {1 + \frac{1}{{s\Delta \tau }}} \right)\left( {1 - {e^{ - s\tau }}} \right),
\end{align}
\begin{align}\label{appeta2}
{\tilde \eta _2} \approx & \frac{{{\tilde \phi _{1}}{e^{ - \frac{s}{2}\tau }}}}{{{\rm{2}}{G_1}}}({1 - {e^{ - \frac{s}{2}\tau }}} )\notag\\
+&\frac{{{\tilde \phi _{2}} + {{\tilde \delta }_2}}}{{{G_2}}}\left[{1 + {e^{ - s\tau }}( {1 - {e^{ - \frac{s}{2}\tau }}})\frac{{1 + \frac{1}{{s\Delta \tau }}}}{{\rm{2}}}}\right]\notag\\
 +& \frac{{{\tilde \phi _{3}}}}{{{G_3}}}\left[ { - 1 + {e^{ - \frac{s}{2}\tau }}({1 - {e^{ - \frac{s}{2}\tau }}})\frac{{1 - \frac{1}{{s\Delta \tau }}}}{{\rm{2}}}}\right]{e^{ - \frac{s}{2}\tau }}\notag\\
 +& {e^{ - \frac{s}{2}\tau }}({1 - {e^{ - \frac{s}{2}\tau }}})\left[ {\frac{{1 + \frac{1}{{s\Delta \tau }}}}{{\rm{2}}}{\tilde h_{2'1}} + \frac{{1 - \frac{1}{{s\Delta \tau }}}}{{\rm{2}}}\left( {{\tilde h_{31'}} - {{\tilde \delta }_1}\left(1 - {e^{ - s\tau }} \right)} \right)} \right]\notag\\
 +& {\tilde h_{2'}} - {\tilde h_{32}} - {{\tilde \delta }_1}{e^{ - s\tau }} - {{\tilde \delta }_3}{e^{ - \frac{s}{2}\tau }} - {{\tilde \delta }_2},
\end{align}
\begin{align}\label{appeta2pie}
{\tilde \eta _{2'}} \approx \frac{{{\tilde \phi _{2}} + {{\tilde \delta }_2}}}{{{G_2}}},
\end{align}
\begin{align}\label{appet3}
{\tilde \eta _3} \approx \frac{{{\tilde \phi _{3}}}}{{{G_3}}},
\end{align}
and
\begin{align}\label{appet3pie}
{\tilde \eta _{3'}} \approx & \frac{{{\tilde \phi _{1}}{e^{ - \frac{s}{2}\tau }}}}{{{\rm{2}}{G_1}}}( {1 - {e^{ - \frac{s}{2}\tau }}})\notag\\
+& \frac{{({\tilde \phi _{2}} + {{\tilde \delta }_2})}}{{{G_2}}}\left[ { - 1 + {e^{ - \frac{s}{2}\tau }}( {1 - {e^{ - \frac{s}{2}\tau }}})\frac{{1 + \frac{1}{{s\Delta \tau }}}}{{\rm{2}}}}\right]{e^{ - \frac{s}{2}\tau }}\notag\\
  +&\frac{{{\tilde \phi _{3}}}}{{{G_3}}}\left[ {1 + {e^{ - s\tau }}( {1 - {e^{ - \frac{s}{2}\tau }}} )\frac{{1 - \frac{1}{{s\Delta \tau }}}}{{\rm{2}}}}\right]\notag\\
 +& {e^{ - \frac{s}{2}\tau }}( {1 - {e^{ - \frac{s}{2}\tau }}} )\left[ {\frac{{1 + \frac{1}{{s\Delta \tau }}}}{{\rm{2}}}{\tilde h_{2'1}} + \frac{{1 - \frac{1}{{s\Delta \tau }}}}{{\rm{2}}}( {{\tilde h_{31'}} - {{\tilde \delta }_1}\left(1 - {e^{ - s\tau }} \right)})} \right]\notag\\
 +& {\tilde h_3} - {\tilde h_{2'3'}} + {{\tilde \delta }_1}{e^{ - \frac{s}{2}\tau }} + {{\tilde \delta }_2}{e^{ - \frac{s}{2}\tau }} + {{\tilde \delta }_3},
\end{align}
where one has introduced
\begin{align}\label{gwhh}
 &{\tilde h_{2'}}{e^{ - \frac{s}{2}\tau}} + {\tilde h_1} \equiv {\tilde h_{2'1}},\notag\\
 &{\tilde h_3}{e^{ - \frac{s}{2}\tau}} + {\tilde h_{1'}} \equiv{\tilde h_{31'}},\notag\\
 &{\tilde h_{2'}}{e^{ - \frac{s}{2}\tau}} + {\tilde h_{3'}} \equiv {\tilde h_{2'3'}},\notag\\
 &{\tilde h_3}{e^{ - \frac{s}{2}\tau}} + {\tilde h_{2}} \equiv{\tilde h_{32}} .
\end{align}

\begin{widetext}
To explicitly address the difference between the standard and arm-locking TDI equations, the coefficients of the science data streams that furnish the TDI equations are given in Tab.~\ref{table1}.
\begin{table}
\caption{\label{table1}The coefficients of science data streams that furnish the standard and arm-locking TDI equations.}
\centering
\newcommand{\tabincell}[2]{\begin{tabular}{@{}#1@{}}#2\end{tabular}}
\begin{ruledtabular}
\renewcommand\arraystretch{2}
\begin{tabular}{|c|c|c|c|c|}
  TDI & science data stream& ${\phi_1}$ & ${\phi_2}$& ${\phi_3}$ \\
 \hline
\multirow{6}*{standard} &${\eta _{1}}$&-1&${{\cal D}_3}$&0\\
 \cline{2-5}
~  &${\eta _{2}}$&0&-1&${{\cal D}_1}$\\
 \cline{2-5}
 ~ &${\eta _{3}}$&${{\cal D}_2}$&0&-1\\
  \cline{2-5}
 ~ &${\eta _{1'}}$&-1&0&${{\cal D}_{2'}}$\\
  \cline{2-5}
 ~ &${\eta _{2'}}$&${{\cal D}_{3'}}$&-1&0\\
  \cline{2-5}
 ~ &${\eta _{3'}}$&0&${{\cal D}_{1'}}$&-1\\
  \hline
 \multirow{6}*{arm locking} &${\tilde \eta _{1}}$&$\frac{1}{2 G_1}(1 - e^{ - s\tau })$&$ - \frac{1}{2 G_2}(1 - \frac{1}{s\Delta \tau })e^{ - \frac{s}{2}\tau }( 1 - e^{ - s\tau })$&$ \frac{1}{2 G_3}(1 - \frac{1}{s\Delta \tau })e^{ - \frac{s}{2}\tau }( 1 - e^{ - s\tau })$\\
  \cline{2-5}
 ~ &${\tilde \eta _{2}}$&$\frac{1}{2G_1}e^{ - \frac{s}{2}\tau }(1 - e^{ - \frac{s}{2}\tau })$&$ \frac{1}{G_2}\left[ 1 + e^{ - s\tau }( 1 - e^{ - \frac{s}{2}\tau })\frac{1 + \frac{1}{s\Delta \tau }}{2} \right]$&$\frac{1}{G_3}\left[ - 1 + e^{ - \frac{s}{2}\tau }( 1 - e^{ - \frac{s}{2}\tau })\frac{1 - \frac{1}{s\Delta \tau }}{2} \right]e^{ - \frac{s}{2}\tau }$\\
  \cline{2-5}
 ~ &${\tilde \eta _{3}}$&0&0&$\frac{1}{G_3}$\\
  \cline{2-5}
 ~ &${\tilde \eta _{1'}}$&$\frac{1}{2 G_1}(1 - e^{ - s\tau })$&$\frac{1}{2 G_2}(1 + \frac{1}{s\Delta \tau })e^{ - \frac{s}{2}\tau }( 1 - e^{ - s\tau })$&$ -\frac{1}{2 G_3}(1 + \frac{1}{s\Delta \tau })e^{ - \frac{s}{2}\tau }( 1 - e^{ - s\tau })$\\
  \cline{2-5}
 ~ &${\tilde \eta _{2'}}$&0&$\frac{1}{G_2}$&0\\
  \cline{2-5}
 ~ &${\tilde \eta _{3'}}$&$\frac{1}{2G_1}e^{ - \frac{s}{2}\tau }(1 - e^{ - \frac{s}{2}\tau })$&$ \frac{1}{G_2}\left[ - 1 + e^{ - \frac{s}{2}\tau }(1 - e^{ - \frac{s}{2}\tau } )\frac{1 + \frac{1}{s\Delta \tau }}{2} \right]e^{ - \frac{s}{2}\tau }$&$ \frac{1}{G_3}\left[1 + e^{ - \frac{s}{2}\tau }(1 - e^{ - \frac{s}{2}\tau } )\frac{1 - \frac{1}{s\Delta \tau }}{2} \right]$\\
\end{tabular}
\end{ruledtabular}
\end{table}
\end{widetext}

Now, we are in the position to explore various TDI solutions in the following subsections Sec.~\ref{section5.2} to Sec.~\ref{section5.4}.

\subsection{The Michelson combination}\label{section5.2}

Similar to the counterpart of conventional TDI solution, the Michelson combinations depend on four data streams, $\tilde \eta_1,\tilde \eta_{1'},\tilde \eta_{2'}$, and $\tilde \eta_{3}$. 
It is readily verified that
\begin{align}\label{mins}
{\tilde \eta _1} - {\tilde\eta _{1'}} \simeq& \left[ {\frac{{{{\tilde \phi }_3}}}{{{G_3}}} - \frac{{({{\tilde \phi }_2} + {{\tilde \delta }_2})}}{{{G_2}}}} \right]{e^{ - \frac{s}{2}\tau }}\left( {1 - {e^{ - s\tau }}} \right)\notag\\
 +& \left[ {{{\tilde h}_{31'}} - {{\tilde \delta }_1}\left( {1 - {e^{ - s\tau }}} \right) - {{\tilde h}_{2'1}}} \right]\left( {1 - {e^{ - s\tau }}} \right),
\end{align}
and
\begin{align}\label{lowsignal}
{\tilde \eta _3} - {\tilde \eta _{2'}} \simeq \frac{{{\tilde \phi _{3}}}}{{{G_3}}} - \frac{{{\tilde \phi _{2}} + {{\tilde \delta }_2}}}{{{G_2}}}.
\end{align}

One observes that the pattern of the laser phase noise $\tilde \phi_{2}$ and $\tilde \phi_{3}$ present in Eqs.~\eqref{mins} and~\eqref{lowsignal} are largely identical. 
Therefore, they can be removed by the following linear combination, which gives rise to the Michelson arm-locking TDI solution
\begin{align}\label{micharm}
	{\rm{TDI}}_{Arm-X}\simeq \left( {{\tilde \eta _1} - {\tilde \eta _{1'}}} \right) - \left( {{\tilde \eta _3} - {\tilde \eta _{2'}}} \right){e^{ - \frac{s}{2}\tau }}\left( {1 - {e^{ - s\tau }}} \right) .
\end{align}

Subsequently, one can read off the TDI coefficients by comparing Eq.~\eqref{micharm} against the general form of the TDI solution Eq.~\eqref{tdi}.
We find
\begin{align}\label{coffXX}
\tilde P_1(s) =& 1,\tilde P_2(s) = 0,\tilde P_3(s) =- e^{-\frac{s}{2}\tau}(1-e^{-s\tau}),\notag\\
\tilde P_{1'}(s) =&-1,\tilde P_{2'}(s) = e^{-\frac{s}{2}\tau}(1-e^{-s\tau}),\tilde P_{3'}(s) = 0.
\end{align}
It is noted that the TDI solution Eqs.~\eqref{coffXX} is different from Eqs.~\eqref{tdipoly}, by taking into consideration the relation Eq.~\eqref{FourierXLaplace} and the approximation Eq.~\eqref{tauapp}.

By substituting the observables given by Eqs.~\eqref{finallyeta1} to~\eqref{appet3pie} and recalling the definitions given by Eqs.~\eqref{new} and~\eqref{gwhh},  the TDI solution Eq.~\eqref{micharm} can be evaluated further to give
\begin{align}\label{micharmsim}
	{\rm{TDI}}_{Arm-X}= \left[ {{{\tilde h}_{31'}} - {{\tilde \delta }_1}\left( {1 - {e^{ - s\tau }}} \right) - {{\tilde h}_{2'1}}} \right]\left( {1 - {e^{ - s\tau }}} \right) .
\end{align}
In particular, the GW signals and floor noise, i.e., the test mass noise and shot noise, are given by
\begin{align}\label{michgwsignal}
{\rm{TDI}}^h_{Arm-X}=\left( { - {h_1}{\rm{ + }}{h_3}{e^{ - \frac{s}{2}\tau }} + {h_{1'}} - {h_{2'}}{e^{ - \frac{s}{2}\tau }}} \right)\left( {1 - {e^{ - s\tau }}} \right),
\end{align}
and
\begin{align}\label{michgwfloornoise}
{\rm{TDI}}_{Arm - X}^{\delta  + {N^S}} = \begin{array}{c}
\left[ \begin{array}{l}
 - {\delta _1}\left( {1 + {e^{ - s\tau }}} \right) + 2{\delta _3}{e^{ - \frac{s}{2}\tau }}\\
 + {\delta _{1'}}\left( {1 + {e^{ - s\tau }}} \right) - 2{\delta _{2'}}{e^{ - \frac{s}{2}\tau }}
\end{array} \right]\left( {1 - {e^{ - s\tau }}} \right)\\
 + \left(-N_1^S+N_3^S{e^{ - \frac{s}{2}\tau }} + N_{1'}^S - N_{2'}^S{e^{ - \frac{s}{2}\tau }} \right)\left( {1 - {e^{ - s\tau }}} \right)
\end{array}.
\end{align}
It is worth noting that in terms of the explicit forms defined by Eq.~\eqref{gwhh}, it is readily verified that both the GW signal and floor noise possess the same forms as their counterparts given by Eqs.~\eqref{tdih} and~\eqref{tdidetlashot} where one substitutes the TDI coefficients Eq.~\eqref{tdipoly} and the approximation Eq.~\eqref{tauapp}. 
In other words, the residual noise PSD and response function can be calculated using Eqs.~\eqref{xben} and~\eqref{responX}, where the coefficients are defined by those of the standard first-generation Michelson TDI combination, namely, Eq.~\eqref{tdipoly}.
Further details of the calculations can be found in Appendix~\ref{apptidal}.
We show the resultant sensitivity curve of the arm-locking TDI combination, governed by Eq.~\eqref{requ}, in Fig.~\ref{fig7}.
According to the above discussions, it is identical to its standard first-generation counterpart without arm locking.
We will come back to further discussions regarding such apparent coincidence in the last section of the paper.

\subsection{The Monitor and Beacon combinations}\label{section5.3}

A similar approach can be used to derive the Monitor and Beacon combinations.
In the case of Monitor combination, the master spacecraft (SC1) acts as a sole recipient that only receives signals from the others.
In other words, the TDI combination depends only on four data streams $\eta_1,\eta_{1'},\eta_2$, and $\eta_{3'}$.
The difference between $\eta _2$ and $\eta _{3'}$ gives
\begin{align}\label{3arm2}
&\left( {{\tilde \eta _2} - {\tilde \eta _{3'}}} \right) \simeq \left( {\frac{{{{\tilde \phi }_2} + {{\tilde \delta }_2}}}{{{G_2}}} - \frac{{{{\tilde \phi }_3}}}{{{G_3}}}} \right)\left( {1 + {e^{ - \frac{s}{2}\tau }}} \right)\notag\\
 &- \left[ \begin{array}{l}
{{\tilde h}_3} - {{\tilde h}_{2'}} + {{\tilde h}_{32}} - {{\tilde h}_{2'3'}} + {e^{ - \frac{s}{2}\tau }}\left( {{e^{ - \frac{s}{2}\tau }} + 1} \right){{\tilde \delta }_1}\\
 + \left( {{e^{ - \frac{s}{2}\tau }} + 1} \right){{\tilde \delta }_2} + \left( {{e^{ - \frac{s}{2}\tau }} + 1} \right){{\tilde \delta }_3}
\end{array} \right].
\end{align}
By canceling out relevant laser phase noise in Eq.~\eqref{mins} and Eq.~\eqref{3arm2}, one obtains the desired combination
\begin{align}\label{tdicom}
&{\rm{TDI}}_{Arm-E} \simeq  \left( {{\tilde \eta _1} - {\tilde \eta _{1'}}} \right) + \left( {{e^{ - \frac{s}{2}\tau }} - {e^{ - s\tau }}} \right)\left( {{\tilde \eta _2} - {\tilde \eta _{3'}}} \right) .
\end{align}

This indicates the following Monitor arm-locking TDI solution 
\begin{align}\label{coffEE}
\tilde P_1(s) =& 1,\tilde P_2(s) =(e^{-\frac{s}{2}\tau}-e^{-s\tau}),\tilde P_3(s) = 0,\notag\\
\tilde P_{1'}(s) =& -1,\tilde P_{2'}(s) = 0,\tilde P_{3'}(s) =-(e^{-\frac{s}{2}\tau}-e^{-s\tau}) ,
\end{align}
which is distinct from the standard first-generation Monitor combination.

One may rewrite Eq.~\eqref{tdicom} explicitly in terms of the GW signal and residual noise as
\begin{align}\label{tdicomssim}
{\rm{TDI}}_{Arm-E} =&  \!-\! {e^{ - \frac{s}{2}\tau }}\left( {1 \!-\! {e^{ - \frac{s}{2}\tau }}} \right)({{\tilde h}_2}\! -\!{{\tilde h}_{3'}})
\!+\!\left( {1 \!-\! {e^{ - s\tau }}} \right)( {{\tilde h}_{1'}}\!-\!{{\tilde h}_1})\notag\\
	-& {{\tilde \delta }_1}\left( {1 - {e^{ - s\tau }}} \right) - {e^{ - \frac{s}{2}\tau }}\left( {1 - {e^{ - s\tau }}} \right)\left( {{{\tilde \delta }_2} + {{\tilde \delta }_3}} \right) .
\end{align}
Again, they are in accordance with Eqs.~\eqref{tdih} and~\eqref{tdidetlashot} in terms of the coefficients of the standard first-generation Monitor combination.
To be specific, we have
\begin{align}\label{monitorgwsignal}
{\rm{TDI}}_{Arm - E}^h =  - \left( {{e^{ - \frac{s}{2}\tau }} - {e^{ - s\tau }}} \right)\left( {{h_2} - {h_{3'}}} \right) + \left( {1 - {e^{ - s\tau }}} \right)\left( {{h_{1'}} - {h_1}} \right),
\end{align}
and
\begin{align}\label{monitorfloornoise}
{\rm{TDI}}_{Arm - E}^{\delta  + {N^S}} =& \left( {1 - {e^{ - s\tau }}} \right)\left( {{\delta _{1'}} - {\delta _1}} \right)\notag\\
 -& \left( {1 - {e^{ - s\tau }}} \right){e^{ - \frac{s}{2}\tau }}\left( {{\delta _{2'}} - {\delta _3}} \right)\notag\\
 +& \left( { - {e^{ - \frac{s}{2}\tau }} + 2{e^{ - s\tau }} - {e^{ - \frac{{3s}}{2}\tau }}} \right)\left( {{\delta _2} - {\delta _{3'}}} \right)\notag\\
 -& {e^{ - \frac{s}{2}\tau }}\left( {1 - {e^{ - \frac{s}{2}\tau }}} \right)\left( {N_2^S - N_{3'}^S} \right) + \left( {1 - {e^{ - s\tau }}} \right)\left( {N_{1'}^S - N_1^S} \right) ,
\end{align}
from which the response function and residual noise PSD can be derived as
\begin{align}\label{resmonitor}
R{(u)_{Arm-E}} =& \frac{1}{6}{\sin ^2}\frac{u}{2}\{ 20 + 4\cos u + 168\left( {1 + \cos u} \right){\rm{Ci}}u \notag\\
-& 240\left( {1 + \cos u} \right){\rm{Ci}}2u\notag\\
 -& \frac{3\left(27\sin u - 9\sin 2u + \sin 3u \right)}{u}\notag\\
  -& \frac{3\left(11 - 15\cos u + 11\cos 2u + 5\cos 3u \right)}{u^2}\notag\\
 +& \frac{3\left(  - 25\sin u + 11\sin 2u + 5\sin 3u\right)}{u^3}\notag\\
 +& 24[3\left(1 + \cos u \right){\rm{Ci}}3u + \left(1 + \cos u \right)\log \frac{1024}{27}\notag\\
 + &3\sin u\left( {\rm{Si}}u - 2{\rm{Si}}2u + {\rm{Si}}3u \right)]\},
\end{align}
and
\begin{align}\label{noisemonitor}
N(u)_{Arm-E} = 8{\sin ^2}\frac{u}{2}\left[ {2\left( {3 + \cos u} \right)\frac{{s_a^2{L^2}}}{{{u^2}{c^4}}} + \left( {3 + 2\cos u} \right)\frac{{{u^2}s_x^2}}{{{L^2}}}} \right].
\end{align}
The resultant sensitivity curve defined by Eq.~\eqref{requ} is also shown in Fig.~\ref{fig7}.

In the case of the Beacon combination, the master spacecraft (SC1) acts as a sole emitter that only transmits signals to the others.
The Beacon is subsequently found to be
\begin{align}\label{beaarm}
{\rm{TDI}}_{Arm-P}\simeq \left( {{\tilde \eta _3} - {\tilde \eta _{2'}}} \right)\left( {1 - {e^{ - s\tau }}} \right){e^{ - \frac{s}{2}\tau }} + \left( {{e^{ - \frac{s}{2}\tau }} - {e^{ - s\tau }}} \right)\left( {{\tilde \eta _2} - {\tilde \eta _{3'}}} \right) ,
\end{align}
which corresponds to the following TDI coefficients
\begin{align}\label{coffbeacon}
\tilde P_1(s) =&0,\tilde P_2(s) = (e^{-\frac{s}{2}\tau}-e^{-s\tau}),\tilde P_3(s) = (e^{-\frac{s}{2}\tau}-e^{-\frac{3s}{2}\tau}),\notag\\
\tilde P_{1'}(s) =& 0,\tilde P_{2'}(s) = -(e^{-\frac{s}{2}\tau}-e^{-\frac{3s}{2}\tau}),\tilde P_{3'}(s) = -(e^{-\frac{s}{2}\tau}-e^{-s\tau}).
\end{align}

Similarly, the GW signal and residual noise are given by
\begin{align}\label{beaharmsim}
{\rm{TDI}}_{Arm-P} = - \left( e^{ - \frac{s}{2}\tau } - e^{ - s\tau} \right)\left[ \begin{array}{l}
 	{\tilde h_3} - {\tilde h_{2'}} + {\tilde h_{32}} - {\tilde h_{2'3'}} \\
 	+ {e^{ - \frac{s}{2}\tau }}({e^{ - \frac{s}{2}\tau }} + 1){{\tilde \delta }_1}\\
 	+ ({e^{ - \frac{s}{2}\tau }} + 1){{\tilde \delta }_2} + ({e^{ - \frac{s}{2}\tau }} + 1){{\tilde \delta }_3}
 \end{array} \right] .
\end{align}
By rewriting the above expression as
\begin{align}\label{beacongwsignal}
{\rm{TDI}}_{Arm - P}^h =&  - \left( {{e^{ - \frac{s}{2}\tau }} - {e^{ - \frac{{3s}}{2}\tau }}} \right)\left( {{h_3} - {h_{2'}}} \right) \notag\\
-& \left( {{e^{ - \frac{s}{2}\tau }} - {e^{ - s\tau }}} \right)\left( {{h_2} - {h_{3'}}} \right),
\end{align}
and
\begin{align}\label{beaconfloornoise}
{\rm{TDI}}_{Arm - P}^{\delta  + {N^S}} =&  - {e^{ - s\tau }}\left( {1 - {e^{ - s\tau }}} \right)\left( {{\delta _{1'}} - {\delta _1}} \right)\notag\\
 -& {e^{ - \frac{s}{2}\tau }}\left( {1 - {e^{ - s\tau }}} \right)\left( {{\delta _3} - {\delta _{2'}}} \right)\notag\\
 +& \left( { - {e^{ - \frac{s}{2}\tau }} + 2{e^{ - s\tau }} - {e^{ - \frac{{3s}}{2}\tau }}} \right)\left( {{\delta _2} - {\delta _{3'}}} \right)\notag\\
 -& \left( {{e^{ - \frac{s}{2}\tau }} - {e^{ - \frac{{3s}}{2}\tau }}} \right)\left( {N_3^S - N_{2'}^S} \right)\notag\\
 -& \left( {{e^{ - \frac{s}{2}\tau }} - {e^{ - s\tau }}} \right)\left( {N_2^S - N_{3'}^S} \right),
\end{align}
their consistency with the standard counterparts are readily shown.
The above results give rise to the following forms of the response function and residual noise PSD
\begin{align}\label{resposenarmbeacon}
R(u)_{Arm - P}= R(u)_{Arm - E},
\end{align}
and
\begin{align}\label{noisearmbeacon}
N{\left( u \right)_{Arm - P}} = N{\left( u \right)_{Arm - E}}.
\end{align}
The resulting sensitivity curve of the arm-locking TDI Beacon combination is also shown in Fig.~\ref{fig7}.

\subsection{Fully Symmetric Sagnac combination}\label{section5.4}

The fully symmetric Sagnac combination is constructed by canceling out laser noise through the totally symmetric combination of the TDI observables.
Here, a similar strategy can be carried out. 
By appropriately normalizing the laser phase noise, one finds 
\begin{align}\label{symmearm}
{\rm{TDI_{Arm - \zeta}} } \simeq&{e^{ - \frac{s}{2}\tau }}\left[ {-\left( {{\tilde \eta _2} - {\tilde \eta _{3'}}} \right) + \left( {{\tilde \eta _3} - {\tilde \eta _{2'}}} \right)} \right]\notag\\
-& \frac{e^{ - \frac{s}{2}\tau }}{\left( {1 - {e^{ - s\tau }}} \right)}\left( {\tilde \eta _1} - {\tilde \eta _{1'}} \right),
\end{align}
whose coefficients gives the TDI solution
\begin{align}\label{coffzeta}
\tilde P_1(s) =& -\frac{e^{ - \frac{s}{2}\tau }}{\left( {1 - {e^{ - s\tau }}} \right)},\tilde P_2(s) = e^{-\frac{s}{2}\tau},\tilde P_3(s) = e^{-\frac{s}{2}\tau},\notag\\
\tilde P_{1'}(s) =& \frac{e^{ - \frac{s}{2}\tau }}{\left( {1 - {e^{ - s\tau }}} \right)},\tilde P_{2'}(s) = -e^{-\frac{s}{2}\tau},\tilde P_{3'}(s) =-e^{-\frac{s}{2}\tau} ,
\end{align}
which is different from the standard fully symmetric sagnac combination.

\begin{figure}[!t]
\includegraphics[width=0.40\textwidth]{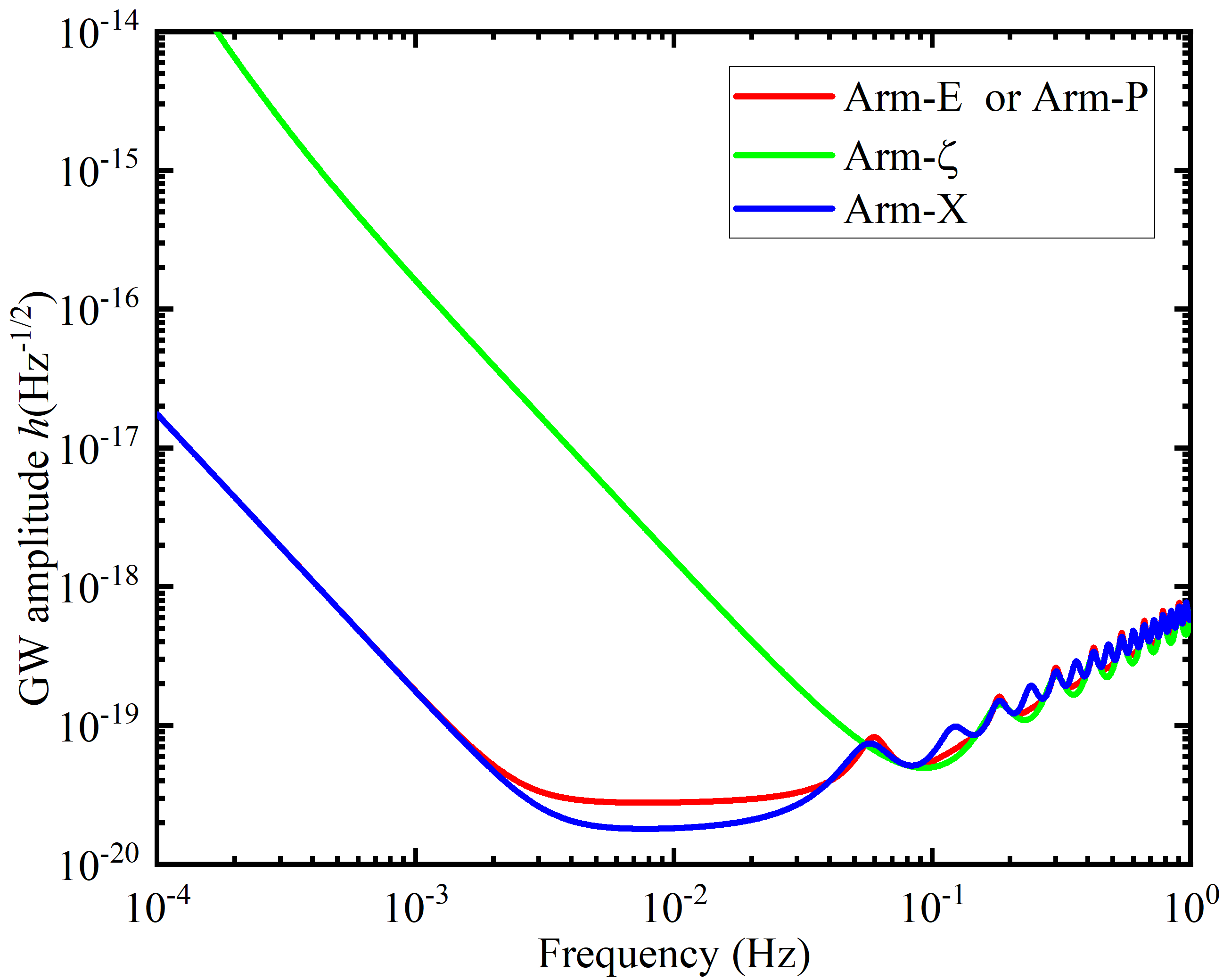}
\caption{\label{fig7}
The resulting sensitivity curves of various arm-locking TDI combinations. The calculations were carried out using the parameters of the LISA detector. The armlength~ $L=2.5\times10^6 {\rm{km}},s_a = 3 \times {10^{ - 15}}\frac{\rm {m/s^2}}{\sqrt {\rm{Hz}} }, s_x = 15 \times {10^{ - 12}}\frac{\rm{m}}{{\sqrt {\rm{Hz}} }}$.}
\end{figure}

Using the specific forms of the observables, Eq.~\eqref{symmearm} gives 
\begin{align}\label{symmearmtdi}
{\rm{TDI}}_{Arm - \zeta } = - e^{ - \frac{s}{2}\tau }\left[ \begin{array}{l}
	{\tilde h_{1'}} + {\tilde h}_{2'}+{\tilde h}_{3'}  - \left( {\tilde h}_1 +{\tilde h}_2 + {\tilde h}_3\right)\\
	- (1 + e^{ - \frac{s}{2}\tau }){\tilde \delta }_1 - (1 + e^{ - \frac{s}{2}\tau }){\tilde \delta }_2 - (1 + {e^{ - \frac{s}{2}\tau }}){{\tilde \delta }_3}
\end{array} \right].
\end{align}
In particular, the GW signals and floor noise are given by
\begin{align}\label{zetagwsignal}
TDI_{Arm - \zeta }^h = {e^{ - \frac{s}{2}\tau }}\left[ {\left( {{h_1} + {h_2} + {h_3}} \right) - ({h_{1'}} + {h_{2'}} + {h_{3'}})} \right],
\end{align}
and
\begin{align}\label{zetafloornoise}
TDI_{Arm - \zeta }^{\delta  + {N^S}} =& \left( {{e^{ - \frac{s}{2}\tau }} - {e^{ - s\tau }}} \right)\left[ {\left( {{\delta _1} + {\delta _2} + {\delta _3}} \right) - \left( {{\delta _{1'}} + {\delta _{2'}} + {\delta _{3'}}} \right)} \right]\notag\\
 +& {e^{ - \frac{s}{2}\tau }}\left[ {\left( {N_1^S + N_2^S + N_3^S} \right) - (N_{1'}^S + N_{2'}^S + N_{3'}^S)} \right].
\end{align}
Subsequently, the response function and residual noise PSD are found to be
\begin{align}\label{resposenarmzeta}
R\left( u \right)_{{Arm -\zeta}} =& 9\cos u\log \frac{4}{3} + 6\log 2 + {\sin ^2}\frac{u}{2}\notag\\
 -& \frac{{3\left( { - 1 + \cos u} \right)\sin u}}{{4u}} + \frac{{15\cos u{{\sin }^2}\frac{u}{2}}}{{2{u^2}}}\notag\\
 -& \frac{{15{{\sin }^2}\frac{u}{2}\sin u}}{{2{u^3}}} + 9\sin u\left( {{\rm{Si}}u - 2{\rm{Si}}2u + {\rm{Si}}3u} \right)\notag\\
 +& 6({\rm{Ci}}u - {\rm{Ci}}2u) + 9\cos u\left( {{\rm{Ci}}u - 2{\rm{Ci}}2u + {\rm{Ci}}3u} \right),
\end{align}
and
\begin{align}\label{noisearmzeta}
N{\left( u \right)_{{Arm -\zeta}}}= - 12\left( { - 1 + \cos u} \right)\frac{{s_a^2{L^2}}}{{{u^2}{c^4}}} + 6\frac{{{u^2}s_x^2}}{{{L^2}}} .
\end{align}
The resulting sensitivity curve of the arm-locking TDI fully symmetric Sagnac combination is shown in Fig.~\ref{fig7} on top of the others.

\section{Further discussions and concluding remarks}\label{section6}

In this work, we investigate the specific forms of the data streams of the dual-arm locking scheme in conjunction with the TDI algorithm.
Moreover, the observables $\tilde\eta_i$ were constructed from these data streams as the output of the arm locking feedback routing.
Subsequently, the resultant TDI solutions were constructed explicitly in terms of the above observables for the Michelson, Monitor, Beacon, and fully symmetric Sagnac combinations.
The coefficients were identified by comparing with the general form of the TDI combination.
The TDI solutions were found to be distinct from the standard first-generation TDI ones for the above three types of combinations, which was expected due to the deformed laser phase noise via arm locking.
On the other hand, the resultant GW signals, floor noise PSDs, and sensitivity curves were shown to be identical to their conventional counterparts.

The arm locking technique was introduced in the literature to suppress the laser phase noise through optical feedback routing.
When used in conjunction with the TDI algorithm, it is generally understood that arm locking may provide an additional layer of safety to guarantee the successful execution of the former.
However, since the suppressed laser phase noise is deformed in the output data stream as manifestly shown in Eqs.~\eqref{PHI1} to~\eqref{eta22222}, the explicit forms of the TDI solution, as well as the relationship of their standard counterparts, are not straightforward.
To this end, the present study is aimed to explore the properties, particularly the explicit forms of the arm-locking TDI combinations.

Regarding the spacecraft motion noise, we proposed a real-time AOM scheme to eliminate the noise due to optical bench vibrations.
It is based on the hard-wired AOM to compensate for the noise associated with the optical bench vibrations. 
The idea of canceling out the spacecraft jitter by employing the interferometry between a remote and the local spacecraft regarding the difference between a phase-locking signal and an arm-locking one was first proposed in Ref.~\cite{arm-time-2010}.
The present paper suggested an implementation of the above idea.
It is noted that we did not explicitly consider the clock jitter noise in the present study.
In practice, clock noise can be eliminated as a part of the TDI algorithm by introducing sideband data stream~\cite{tdi-clock4}.
Moreover, the notion of optical frequency comb~\cite{tdi-clock5} in space missions has attracted much interest recently.
We argue that one may synchronize the clock jitter noise to the laser phase noise by employing a optical frequency comb and then suppress it simultaneously with the laser phase noise~\cite{tdi-clock5}.
In this regard, the noise floor can be effectively suppressed, particularly in the lower frequency band constituted by the clock jitter. 

We note that the specific forms of the arm-locking TDI solutions depend on the proposed AOM scheme.
From a mathematical perspective, the hard-wired AOM combines the data stream to eliminate the optical bench vibration.
This is equivalent to the algebraic approach for canceling the optical bench noise in the TDI algorithm.
In the case of the TDI algorithm, algebraically, one removes the twelve variables associated with all possible projections of the optical vibrations $\vec{n}_{j (j')}\vec{\Delta}_{i (i')}$ by the twelve equations associated with the test-mass and reference data streams.
As a result, one arrives at three equations in terms of six variables $\eta_{i (i')}$ Eqs.~\eqref{postpre} and~\eqref{postpre1}~\cite{tdi-03}.
For the present study, the algebraic combination is rather different, but one also results in six variables $\tilde\eta_{i (i')}$ Eqs.~\eqref{newetai} and~\eqref{newetaipie}.
On the practical side, as shown in Fig.~\ref{fig5}, the AOM is implemented {\it before} any measurement to modify the input signals of the arm-locking technique.
Therefore, one does not need to elaborate an additional scheme later to eliminate $\vec{n}_{i'}\vec{\Delta}_i$.
On the contrary, if AOM were not implemented, the terms that involve $\vec{\Delta}_{i (i')}$ will be deformed and present in Eqs.~\eqref{newetai} and~\eqref{newetaipie}, and subsequently, Eqs.~\eqref{finallyeta1} to~\eqref{appet3pie}.
Specifically, one will face the complete set of eighteen data streams, which involve the optical bench noise in its modified form. 
It is still possible, but not straightforward, to eliminate them algebraically using the complete set of data streams alongside the TDI approach.
In other words, for any experimental facility which does not possess an AOM, the arm-locking TDI solution must be modified.
Such solutions lie outside of the scope of the present work.

Last but not least, we comment on the apparent {\it coincidence} in the forms of response functions, residual noise PSDs, and subsequently, the sensitivity curves for the four types of TDI combinations studied in this work.
Since the data streams from the arm-locking routing have mixed up with the specific configuration of the arm-locking scheme, it is intuitive that the resultant variables $\tilde\eta_i$ possess more complicated forms, where the laser phase noise is suppressed while deformed.
As expected, the coefficients of the TDI combinations encountered in the present work differ from the original TDI approach.
The resultant GW signal and remaining noise, on the other hand, are also deformed in the arm-locking TDI variables.
This is consistent with the observation that the one-to-one mapping between GW stress and laser noise in the standard variables Eqs.~\eqref{postpre} and~\eqref{postpre1} is no longer valid in the arm-locking TDI observables.
Subsequently, the response function and residual noise PSDs are furnished by summating these deformed magnitudes, which are further scaled to the above TDI coefficients.
For some reason obscure to us, these two factors cancel out precisely.
Moreover, it seems that the method employed in the present study cannot be extended straightforwardly to Relay and Sagnac combinations.
We relegate further discussions concerning this point to Appendix~\ref{apptidal} of the paper.
Nonetheless, a more general approach based on the computational algebraic geometry to exhaustively enumerate the solutions would be desirable.
We plan to explore these relevant topics further in future work.

\section*{Acknowledgements}

This work is supported by the National Natural Science Foundation of China (Grant No.11925503), the Postdoctoral Science Foundation of China (Grant No.2022M711259), Guangdong Major project of Basic and Applied Basic Research (Grant No.2019B030302001), Natural Science Foundation of Hubei Province(2021CFB019), and the Fundamental Research Funds for the Central Universities, HUST: 2172019kfyRCPY029.
We also gratefully acknowledge the financial support from Brazilian agencies Funda\c{c}\~ao de Amparo \`a Pesquisa do Estado de S\~ao Paulo (FAPESP), Funda\c{c}\~ao de Amparo \`a Pesquisa do Estado do Rio de Janeiro (FAPERJ), Conselho Nacional de Desenvolvimento Cient\'{\i}fico e Tecnol\'ogico (CNPq), Coordena\c{c}\~ao de Aperfei\c{c}oamento de Pessoal de N\'ivel Superior (CAPES).

\appendix

\section{Justification of the use of first-generation TDI algorithm and noise in arm-locking TDI}\label{appd2}

In the present paper, the arm-locking technique has been used in conjunction with the first-generation TDI algorithm.
As discussed by a few authors, this can be justified since the presence of the arm locking provides an additional 2-4 orders of magnitude in laser phase noise suppression.
Subsequently, the use of first-generation TDI suffices for the required sensitivity leading to possible GW detection.

To be more specific, we make the following estimations following~\cite{armnewdesign-2022,arm-the-2022}.
The first-generation TDI considers static arm lengths.
By assuming the worst-case scenario that the relative velocity between spacecraft is
\begin{equation}
\Delta v \sim 10 \mathrm{ms}^{-1},\nonumber
\end{equation}
which leads to an armlength mismatch of
\begin{equation}
\Delta L_{\mathrm{TDI1.0}} \lesssim 166 \mathrm{m} .\nonumber
\end{equation}
As a result, the upper bound of the laser frequency noise is estimated to be~\cite{armnewdesign-2022,arm-the-2022}
\begin{equation}
\tilde{\delta \nu}_{\mathrm{TDI1.0}}(f) \sim 1.7\times \sqrt{1+\left(\frac{f_m}{f}\right)^4}\frac{\mathrm{Hz}}{\sqrt{\mathrm{Hz}}} ,\nonumber
\end{equation}
where $f_m$ is the optimal frequency where the magnitudes of the acceleration and shot noises coincide.
For the second-generation TDI, the arm length mismatch is caused by second-order terms.
If one assumes the eventual arm length mismatch to be
\begin{equation}
\Delta L_{\mathrm{TDI2.0}} \lesssim 1 \mathrm{m} ,\nonumber
\end{equation}
we have
\begin{equation}
\tilde{\delta \nu}_{\mathrm{TDI2.0}}(f) \sim 282\times \sqrt{1+\left(\frac{f_m}{f}\right)^4}\frac{\mathrm{Hz}}{\sqrt{\mathrm{Hz}}} .\nonumber
\end{equation}

On the other hand, the arm-locking technique typically provides a suppression in the laser frequency noise of 2-4 orders of magnitude~\cite{armnewdesign-2022,arm-the-2022, arm-moddual-2009}.
Compared to the difference between the first and second-generation TDI, which is approximately two orders of magnitude, it is competent for the task.
Relevant calculations have been carried out in Refs.~\cite{armnewdesign-2022}, which is the point of departure of the present work.

In this study, we have not explicitly considered the additional noise potentially introduced due to the implementation of AOM.
In principle, as a piece of physical equipment, it inevitably causes uncertainty, among which the primary source is related to the driving signals. 
To give an estimation, assume that the frequency shift caused by AOM is 
\begin{equation}
f_{\rm{AOM}}\sim 100\mathrm{MHz}, \notag
\end{equation}
and additional noise can be estimated as~\cite{AOM-noise}
\begin{equation}
f_{\rm{AOM}}\times q ,\notag
\end{equation}
where
\begin{equation}
q=8.2\times 10^{-14}\sqrt{\frac{\mathrm{Hz}}{f}}\frac{1}{\sqrt{\mathrm{Hz}}} \notag
\end{equation}
is the fractional frequency fluctuations of the ultra-stable oscillator. 
Subsequently, by comparing with the laser frequency noise, which is 
\begin{equation}
\tilde{\delta\nu}_{\mathrm{PDH}}=30\times \sqrt{1+\left(\frac{f_m}{f}\right)^4}\frac{\mathrm{Hz}}{\sqrt{\mathrm{Hz}}},\nonumber
\end{equation} 
where $f_m\sim 2.8 \mathrm{ mHz}$,
it is apparent that the additional noise due to AOM can be neglected.

\section{Derivations of the GW response function and floor noise PSD for the arm locking TDI combination}\label{apptidal}

In this Appendix, by taking the Michelson combination as an example, we give a detailed account of the derivation of Eqs.~\eqref{micharmsim} to~\eqref{michgwfloornoise} in the main text.
The arm-locking Michelson combination is given by Eqs.~\eqref{micharm}. 
\begin{equation}
{\rm{TDI}}_{Arm - X} \approx \left( {{{\tilde \eta }_{\rm{1}}} - {{\tilde \eta }_{{\rm{1'}}}}} \right) - \left( {{{\tilde \eta }_3} - {{\tilde \eta }_{2'}}} \right){e^{ - \frac{s}{2}\tau }}\left( {1 - {e^{ - s\tau }}} \right),\nonumber
\end{equation}
By substituting the definitions of the observables given by Eqs.~\eqref{finallyeta1} to~\eqref{gwhh}, Eq.~\eqref{micharm} gives
\begin{align}
 {\rm{TDI}}_{Arm - X}=& \left[ {{{\tilde h}_{{\rm{31'}}}} - {{\tilde \delta }_1}\left( {1 - {e^{ - s\tau }}} \right) - {{\tilde h}_{{\rm{2'1}}}}} \right]\left( {1 - {e^{ - s\tau }}} \right)\notag\\
 =& \left[ {{{\tilde h}_{\rm{3}}}{e^{ - \frac{s}{2}\tau }} + {{\tilde h}_{{\rm{1'}}}} - {{\tilde \delta }_1}\left( {1 - {e^{ - s\tau }}} \right) - {{\tilde h}_{{\rm{2'}}}}{e^{ - \frac{s}{2}\tau }} - {{\tilde h}_{\rm{1}}}} \right]\left( {1 - {e^{ - s\tau }}} \right).\notag
 \end{align}
One further takes into account the definitions of the GW signal and noise, namely, Eq.~\eqref{new}. The above expression can be rewritten as
 \begin{align}
 {\rm{TDI}}_{Arm - X}= &\left[ \begin{array}{l}
\left( {2{\delta _3} + N_3^S + {h_{\rm{3}}}} \right){e^{ - \frac{s}{2}\tau }} \\
+ (2{\delta _{1'}} + N_{1'}^S + {h_{{\rm{1'}}}})\\
 - \left( {{\delta _{1'}} - {\delta _1}} \right)\left( {1 - {e^{ - s\tau }}} \right)\\
 - \left( {2{\delta _{2'}} + N_{2'}^S + {h_{{\rm{2'}}}}} \right){e^{ - \frac{s}{2}\tau }}\notag\\
 - \left( {2{\delta _1} + N_1^S + {h_{\rm{1}}}} \right)
\end{array} \right]\left( {1 - {e^{ - s\tau }}} \right)\notag\\
=& {\rm{TDI}}_{Arm - X}^h + {\rm{TDI}}_{Arm - X}^{\delta  + {N^S}},\notag
\end{align}
where
\begin{equation}
{\rm{TDI}}_{Arm - X}^h=\left( { - {h_{\rm{1}}} + {h_{\rm{3}}}{e^{ - \frac{s}{2}\tau }} + {h_{{\rm{1'}}}} - {h_{{\rm{2'}}}}{e^{ - \frac{s}{2}\tau }}} \right)\left( {1 - {e^{ - s\tau }}} \right),\notag
\end{equation}
and
\begin{equation}
\begin{aligned}
{\rm{TDI}}_{Arm - X}^{\delta  + {N^S}}=&\left[ \begin{array}{l}
 - {\delta _1}(1 + {e^{ - s\tau }}) + 2{\delta _3}{e^{ - \frac{s}{2}\tau }}\\
 + {\delta _{1'}}(1 + {e^{ - s\tau }}) - 2{\delta _{2'}}{e^{ - \frac{s}{2}\tau }}
\end{array} \right]\left( {1 - {e^{ - s\tau }}} \right)\\
 +& \left( { - N_1^S{\rm{ + }}N_3^S{e^{ - \frac{s}{2}\tau }} + N_{1'}^S - N_{2'}^S{e^{ - \frac{s}{2}\tau }}} \right)\left( {1 - {e^{ - s\tau }}} \right).\notag
 \end{aligned}
\end{equation}
These are the results given by Eqs.~\eqref{micharmsim} to~\eqref{michgwfloornoise}.

Now, one can proceed to show that the above expression for the GW signal and floor noise possess the same forms as their counterparts given by Eqs.~\eqref{tdih} and~\eqref{tdidetlashot}.
Specifically, we substitute the TDI coefficients Eq.~\eqref{tdipoly} into Eqs.~\eqref{tdih} and~\eqref{tdidetlashot}, and find
\begin{equation}
\begin{aligned}
{\rm{TDI}}_{1.0 - X}^h = \left( {1 - {{\cal D}_{33'}}} \right){h_{1'}} + \left( {{{\cal D}_{2'}} - {{\cal D}_{33'2'}}} \right){h_3}\\
 + \left( {{{\cal D}_{2'2}} - 1} \right){h_1} + \left( {{{\cal D}_{2'23}} - {{\cal D}_3}} \right){h_{2'}},\notag
\end{aligned}
\end{equation}
and
\begin{equation}
\begin{aligned}
{\rm{TDI}}_{1.0 - X}^{\delta  + N} =& \left( {1 - {{\cal D}_{33'}}} \right) \delta_{1'} + \left( {{{\cal D}_{2'}} - {{\cal D}_{33'2'}}} \right) {\cal D}_2\delta _{1'}\\
 +& \left( {{{\cal D}_{2'2}} - 1} \right) {\cal D}_3{ \delta }_{2'} + \left( {{{\cal D}_{2'23}} - {{\cal D}_3}} \right) {\delta }_{2'}\\
 +& \left( {{{\cal D}_{2'2}} - 1} \right){\delta }_1 + \left( {{{\cal D}_{2'23}} - {{\cal D}_3}} \right){{\cal D}_{3'}}{\delta }_1\\
 +& \left( {1 - {{\cal D}_{33'}}} \right) {{\cal D}_{2'}}{\delta }_3 + \left( {{{\cal D}_{2'}} - {{\cal D}_{33'2'}}} \right) { \delta }_3\\
 +&\left( {1 - {{\cal D}_{33'}}} \right)N_{1'}^S + \left( {{{\cal D}_{2'}} - {{\cal D}_{33'2'}}} \right)N_3^S \\
 +& \left( {{{\cal D}_{2'2}} - 1} \right)N_1^S + \left( {{{\cal D}_{2'23}} - {{\cal D}_3}} \right)N_{2'}^S.\notag
 \end{aligned}
\end{equation}

By taking the Fourier transform of the above equations and utilizing the approximation Eq.~\eqref{tauapp}.
We arrive at
\begin{equation}
{\rm{TDI}}_{1.0 - X}^h=\left( { - {h_{\rm{1}}} + {h_{\rm{3}}}{e^{ - \frac{s}{2}\tau }} + {h_{{\rm{1'}}}} - {h_{{\rm{2'}}}}{e^{ - \frac{s}{2}\tau }}} \right)\left( {1 - {e^{ - s\tau }}} \right),\notag
\end{equation}
and
\begin{equation}
\begin{aligned}
{\rm{TDI}}_{1.0 - X}^{\delta  + N}=&\left[ \begin{array}{l}
 - {\delta _1}(1 + {e^{ - s\tau }}) + 2{\delta _3}{e^{ - \frac{s}{2}\tau }}\\
 + {\delta _{1'}}(1 + {e^{ - s\tau }}) - 2{\delta _{2'}}{e^{ - \frac{s}{2}\tau }}
\end{array} \right]\left( {1 - {e^{ - s\tau }}} \right)\\
 +& \left( { - N_1^S{\rm{ + }}N_3^S{e^{ - \frac{s}{2}\tau }} + N_{1'}^S - N_{2'}^S{e^{ - \frac{s}{2}\tau }}} \right)\left( {1 - {e^{ - s\tau }}} \right),\notag
 \end{aligned}
\end{equation}
which are manifestly identical to Eqs.~\eqref{micharmsim} to~\eqref{michgwfloornoise}.
Therefore, the residual noise PSD and response function are the same as their first-generation counterparts.

Similar calculations can be carried out for the remaining combinations investigated in this paper.
Apart from the apparent coincidence, we note that the conclusion was drawn based on the following factors:
\begin{itemize}
\item The difference between the armlengths is insignificant.
\item The controllers' gain is significant.
\item The residual laser noise will be below the noise floor.
\end{itemize}
The first factor indicates that one ignores the difference between the time delays in different arms $\Delta\tau$.
In other words, we are dealing with first-generation TDI solutions.
However, as discussed in the main text, $\Delta\tau$ still plays a role owing to its presence in some of the denominators due to the dual-arm locking scheme. 
The second factor implies that expansion will be carried out for $1/G_i$, and the sub-leading terms are ignored.
Lastly, as can be justified, we follow the TDI algorithm's standard practice and assume the residual laser noise is below the noise floor, primarily consisting of test-mass vibration and shot noise.
In this regard, the response function is govered by the terms $h_{i(i')}$ through the definitions Eqs.~\eqref{new} and Eqs.~\eqref{gwhh}.
On the other hand, the noise PSD are determined by the shot noise $N^S_{i(i')}$ and test-mass one $\delta_{i(i')}$, via the respective definitions in Eqs.~\eqref{new} and Eqs.~\eqref{gwhh}.
In particular, the laser noise $\phi_{i(i')}$ will not enter the resultant expressions.
This is because, when expanded in terms of the rate of change of the armlength, the leading contributions are readily canceled out by the very definition of the first-generation TDI solution.
The next-to-leading contributions, which are mostly accounted for by the second-generation TDI, are already below the above noise floor and therefore become irrelevant.

\bibliographystyle{h-physrev}
\bibliography{reference_wang}

\end{document}